\RequirePackage[T1]{fontenc}
\documentclass[journal,unicode,hidelinks]{IEEEtran}
\usepackage{amsmath,amssymb,amsthm}
\usepackage{graphicx}
\usepackage[nocompress]{cite}
\usepackage{algorithm}
\usepackage{algorithmic}
\usepackage{placeins}
\usepackage{stfloats}
\usepackage{orcidlink}
\usepackage{hyperref}
\hypersetup{
    pdfauthor={Muhammad Shoaib Ayub and Insoo Koo},
    pdftitle={VR-Net: Subarray-Level Inter-Cell Coordination
    for Multi-Cell XL-MIMO}
}
\renewcommand{\eqref}[1]{\hyperref[#1]{\textup{(\ref*{#1})}}}
\newcommand{\proofref}[2]{\hyperlink{\getrefbykeydefault{#1}{anchor}{}}{#2}}
\newtheorem{theorem}{Theorem}
\newtheorem{proposition}{Proposition}
\newtheorem{lemma}{Lemma}
\newtheorem{remark}{Remark}
\newtheorem{corollary}{Corollary}
\theoremstyle{definition}

\begin{document}
\title{VR-Net: Subarray-Level Inter-Cell Coordination for
Multi-Cell XL-MIMO under Near-Field Spatial Non-Stationarity}

\author{Muhammad Shoaib Ayub\,\orcidlink{0000-0002-6566-8832},
Rana Muhammad Subhan\,\orcidlink{0009-0006-3453-4707},
Insoo Koo\,\orcidlink{0000-0001-7476-8782},
and Felipe Augusto Pereira de Figueiredo\,\orcidlink{0000-0002-2167-7286}%
\thanks{This research was supported by the ANCHOR Program
through the Ulsan ANCHOR Center, funded by the Ministry
of Education (MOE) and the Ulsan Metropolitan City,
Republic of Korea (2026-ANCHOR-07-001).
This work was partially supported by the xGMobile Project
(XGM-AFCCT-2026-8-27-1) with resources from
EMBRAPII/MCTI (Grant 052/2023 PPI IoT/Manufatura 4.0),
CNPq (302085/2025-4 and 306199/2025-4),
FAPEMIG (APQ-05305-23, PPE-00124-23, and RED-00194-23),
FINEP (No. 1060/2, Contract 01.25.0883.00),
and RNP, with resources from MCTIC
(Grant No. 01245.020548/2021-07),
under the Brazil 6G Project of the Radiocommunication
Reference Center of Inatel, Brazil.
The authors also acknowledge the support of
ABRINT, MODIRUM/GESPI, and NIC.br.}%
\thanks{Muhammad Shoaib Ayub, Rana Muhammad Subhan,
and Insoo Koo are with the Department of Electrical,
Electronic and Computer Engineering, University of Ulsan,
Ulsan, South Korea
(e-mail: muhammad.shoaib@usach.cl;
nsubhan62@mail.ulsan.ac.kr; iskoo@ulsan.ac.kr).}%
\thanks{Felipe Augusto Pereira de Figueiredo is with
the National Institute of Telecommunications (Inatel),
Santa Rita do Sapucaí, MG, Brazil
(e-mail: felipe.figueiredo@inatel.br).}}

\maketitle
\begin{abstract}
Spatial non-stationarity in multi-cell extremely large-scale MIMO (XL-MIMO) confines each user to a visibility region (VR) of each array. VR-Net exploits these regions for subarray-level coordinated beamforming. A base-station (BS)-indexed conflict graph identifies overlapping VRs; gain-based association and a masked multi-cell weighted minimum mean-square error (WMMSE) iteration restrict beam design and channel state information (CSI) uploads to the relevant subarray blocks. We establish conditions for exact per-cell decoupling, derive the expected CSI loads, bound the loss of designing on the VR-gated channel and the loss of the support constraint, and characterize single-edge visibility contrast through a closed-form Fresnel transition width. Simulations use a 3GPP TR 38.901 Release-19-based near-field UMi channel with modular facade arrays and pre-registered criteria. At a 30 dB threshold, VR-Net retains at least 99.6\% of the full-aperture WMMSE sum rate in every tested setting while uploading 10-39\% fewer complex CSI entries. It gains up to 3.5 bit/s/Hz over an all-user adaptation of the multi-cell scheme of Pérez-Adán et al. at moderate power and dense load, and inter-cell coordination adds up to 33\% over per-cell design.
\end{abstract}
\section{Introduction}\label{sec:introduction}
\IEEEPARstart{E}{xtremely} large aperture arrays can place user equipment (UE) in the radiative near field of a base station (BS), with strong channel energy confined to part of its aperture. \hyperlink{cite.3gpp38901}{TR~38.901} Release~19 models this spatial non-stationarity (SNS) through element-wise ray attenuation \cite{3gpp38901}. We define operational link-level visibility regions (VRs) from aggregate subarray energy using the per-ray construction in \cite{xu2026}. Recent work addresses spherical-wave analysis and channel modeling \cite{wang2025edof,xu2026thz,jin2025gbsm}, VR-based grouping, and multi-cell NOMA \cite{zhi2024vr,jayasinghe2025vr}.

In multi-cell networks, overlapping VRs couple transmissions through shared subarrays. Per-cell design ignores this coupling, whereas full-CSI coordination exchanges every UE channel at every BS. VR-Net coordinates beamforming only through the subarray blocks that VRs share, and uploads only those blocks. In a Release-19-based evaluation, this retains at least $99.6\%$ of full-aperture WMMSE with up to $39\%$ fewer CSI entries, and the saving grows with the inter-module gap.

The contributions are as follows.
\begin{itemize}
\item \emph{Multi-cell VR model.} \hyperref[sec:system_model]{Section~\ref*{sec:system_model}} defines the operational VR of each UE at each BS on a customized Release-19-based near-field channel, the multi-cell conflict graph, the victim sets, and the conflict-aware CSI load.
\item \emph{VR-Net.} \hyperref[sec:algorithm]{Section~\ref*{sec:algorithm}} gives a two-timescale algorithm: gain-based association and a masked multi-cell WMMSE iteration, monotone under a safeguard, that exchanges only conflict-relevant blocks.
\item \emph{Analysis.} \hyperref[sec:analysis]{Section~\ref*{sec:analysis}} proves when per-cell beamforming is exact, derives the expected CSI loads, bounds the design-to-true rate gap and the support-constraint loss, and characterizes single-edge visibility contrast and the NLoS power floor.
\item \emph{Evaluation.} \hyperref[sec:results]{Section~\ref*{sec:results}} compares performance on separate drop sets against full-aperture WMMSE and an all-user adaptation of \cite{perezadan2025} with full and VR-gated CSI, and examines power, load, and module-gap effects.
\end{itemize}

\subsection{Related Work}
\label{sec:related}

\emph{Near-field channel models.}
Wang et al.~\cite{wang2025edof} analyze spherical-wave degrees of freedom, and Xing et al.~\cite{xing2025wn} develop a wavenumber-domain model. Xu et al.~\cite{xu2026thz} report THz measurements, while Jin et al.~\cite{jin2025gbsm} model near- and far-field channels. Xu et al.~\cite{xu2026} develop the 6--24~GHz near-field SNS construction adopted in \hyperlink{cite.3gpp38901}{TR~38.901} \cite{3gpp38901}.

\emph{Modular arrays and VR-aware processing.}
Zheng et al.~\cite{zheng2025elaa} and Li et al.~\cite{li2025codebook} design near-field and modular-array codebooks. Meng et al.~\cite{meng2025isac} study modular ISAC beamforming; Lu et al.~\cite{lu2025flex} consider flexible layouts, and Lu et al.~\cite{lu2026compact} analyze compact arrays with mutual coupling. Cao et al.~\cite{cao2025ris} develop RIS-assisted precoding. Che et al.~\cite{che2025pilot} and Zhang et al.~\cite{zhang2026pilot} exploit SNS in pilot design. Zhi et al.~\cite{zhi2024vr} group users by VR overlap; Jayasinghe et al.~\cite{jayasinghe2025vr} study multi-cell XL-MIMO NOMA, and Yilmaz et al.~\cite{yilmaz2026vr} use VR-aware interference suppression. Our focus is the CSI-entry exchange required for coordinated multi-cell downlink beamforming.

\emph{Multi-cell and cell-free beamforming.}
Shi et al.~\cite{shi2011} establish WMMSE-based weighted sum rate design. Wang et al.~\cite{wang2026cf} study near-field cell-free combining, Liao et al.~\cite{liao2026edge} address distributed UE association, and Chen et al.~\cite{chen2025meta} develop learning-based beamforming. P\'erez-Ad\'an et al.~\cite{perezadan2025} use full CSI and a LoS spherical-wave model. Their procedure greedily selects UE--BS pairs, applies multi-cell zero forcing and water-filling, and stops if an added UE lowers sum rate. \hyperref[sec:setup]{Section~\ref*{sec:setup}} specifies our all-user adaptation.

\emph{CSI overhead and research gap.}
Cui et al.~\cite{cui2025ice} and Zhang and Cui~\cite{zhang2026obs} study channel estimation; Lu et al.~\cite{lu2025extrap} address channel extrapolation, and Mobini et al.~\cite{mobini2026fh,mobini2026cluster} develop fronthaul-aware processing. These studies address acquisition or transport costs. VR-Net instead specifies BS-indexed conflict-aware uploads for masked multi-cell WMMSE and analyzes the resulting support and soft-visibility losses.

\section{System Model}
\label{sec:system_model}

\begin{figure*}[t]
\centering
\includegraphics[width=\textwidth]{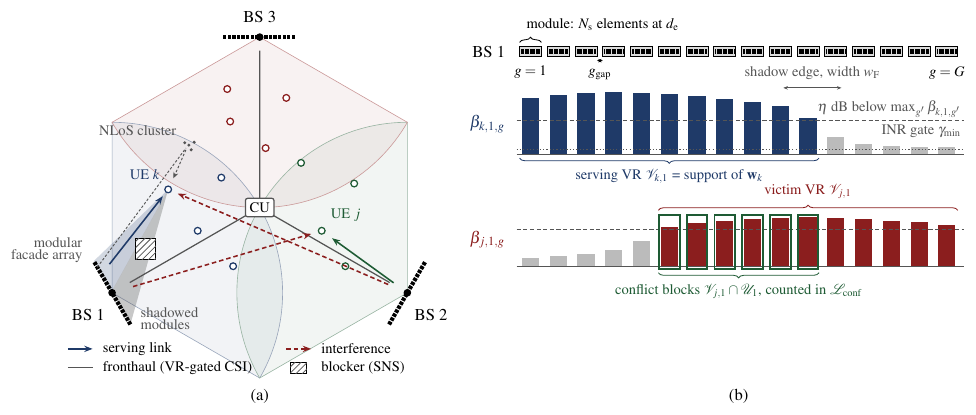}
\caption{System model. (a) Three BSs with modular facade arrays serve UEs in $120^\circ$ sectors; a blocker shadows part of the aperture of BS~1 for UE~$k$, and the BSs exchange only VR-gated CSI with the central unit (CU). (b) Block gains of the served UE~$k$ and of a victim UE~$j$ at BS~1: blocks within $\eta$~dB of the maximum and above the INR gate form the VRs $\mathcal{V}_{k,1}$ and $\mathcal{V}_{j,1}$; only the conflict blocks $\mathcal{V}_{j,1}\cap\mathcal{U}_{1}$ enter $\mathcal{L}_{\mathrm{conf}}$.}
\label{fig:system_model}
\end{figure*}

\subsection{Network Model and Assumptions}
\label{sec:network}

The multi-cell layout, blocker-induced visibility, and VR-gated CSI exchange are illustrated in \hyperref[fig:system_model]{Fig.~\ref*{fig:system_model}}.

Consider $B$ BSs and $K$ single-antenna UEs, with $\mathcal{B}=\{1,\ldots,B\}$, $\mathcal{K}=\{1,\ldots,K\}$ and $\mathcal{G}=\{1,\ldots,G\}$. Each BS has an extremely large aperture array of $N=GN_{\mathrm{s}}$ elements in $G$ modules (subarrays) of $N_{\mathrm{s}}$ elements on one horizontal facade line. The element spacing is $d_{\mathrm{e}}=\lambda/2$ and the inter-module gap is $g_{\mathrm{gap}}$. With $\mathbf{e}_{b}$ the unit vector along the facade, the element positions are
$\mathbf{p}_{b,g,s}
=\mathbf{p}_{b,1,1}
+\big((g-1)(N_{\mathrm{s}}d_{\mathrm{e}}+g_{\mathrm{gap}})+(s-1)d_{\mathrm{e}}\big)
\mathbf{e}_{b}$.
The gap $g_{\mathrm{gap}}$ is extra spacing beyond the contiguous element pitch; $g_{\mathrm{gap}}=0$ gives a contiguous array. UE $k$ is at $\mathbf{q}_{k}\in\mathbb{R}^{3}$. Every UE lies beyond the reactive near field $0.62\sqrt{D_{\mathrm{m}}^{3}/\lambda}$ of every module, with $D_{\mathrm{m}}=(N_{\mathrm{s}}-1)d_{\mathrm{e}}$; no far-field condition is imposed, and a UE may lie inside the Rayleigh distance of one BS and outside that of another. UE $k$ is served by BS $a_{k}$, and $\mathcal{K}_{b}\triangleq\{k:a_{k}=b\}$. Coordinated beamforming (CB) is used: one serving BS per UE, no joint transmission.

We assume: \hypertarget{assumption:A1}{(A1)} downlink, narrowband; \hypertarget{assumption:A2}{(A2)} positions, blockers, VRs and channels fixed over one coordination interval; \hypertarget{assumption:A3}{(A3)} subarray indices local to each BS; \hypertarget{assumption:A4}{(A4)} energy-based visibility (\hyperref[sec:channel]{Section~\ref*{sec:channel}}); \hypertarget{assumption:A5}{(A5)} channel moments conditional on fixed geometry, blockers, VRs and cluster powers, with deterministic LoS and zero-mean uncorrelated ray gains; \hypertarget{assumption:A6}{(A6)} independent unit-power symbols and noise of variance $\sigma_{k}^{2}>0$; \hypertarget{assumption:A7}{(A7)} each BS holds its local channels and uploads blocks to a central unit (CU); \hypertarget{assumption:A8}{(A8)} nonnegative weights $\omega_{k}$; \hypertarget{assumption:A9}{(A9)} the element pattern of \hyperlink{cite.3gpp38901}{TR~38.901}, Table~7.3-1, evaluated per element.

\subsection{Channel Model}
\label{sec:channel}

Let $r_{k,b,g,s}\triangleq\|\mathbf{q}_{k}-\mathbf{p}_{b,g,s}\|_{2}$. The LoS block of module $g$ has the non-uniform spherical-wave entries
\begin{equation}
 \left[\mathbf{h}_{k,b,g}^{\mathrm{LoS}}\right]_{s}
 =
 \left(\frac{d_{k,b,\mathrm{3D}}}{r_{k,b,g,s}}\right)^{21/20}
 A^{E}_{k,b,g,s}
 \sqrt{\chi_{k,b,g,s}^{\mathrm{LoS}}}\,
 e^{-\jmath 2\pi r_{k,b,g,s}/\lambda}.
 \label{eq:nusw_los} \end{equation} Here $d_{k,b,\mathrm{3D}}$ is the reference distance, $A^{E}_{k,b,g,s}$ the element-pattern amplitude, and $\chi_{k,b,g,s}^{\mathrm{LoS}}\in(0,1]$ the SNS power factor.

The NLoS clusters $c\in\mathcal{C}_{k,b}$, with $L_{k,b,c}$ rays each, follow Steps~5--7 of \hyperlink{cite.3gpp38901}{TR~38.901}, Sec.~7.5 (UMi-Street Canyon), with clusters below $-25$~dB of the strongest removed. As in \cite{xu2026}, each cluster is assigned a common spherical-wave source distance $d^{1}_{k,b,c}=s_{k,b,c}(d_{k,b,\mathrm{3D}}+c_{0}\tau_{k,b,c})$, where $\tau_{k,b,c}$ is the cluster delay. The two strongest clusters use $s_{k,b,c}=1$; others draw $s_{k,b,c}\sim\mathrm{Beta}(1.53,1.42)$. The implemented distance is hard-clipped below at $2$~m. The NLoS excess delay $\Delta\tau$ of \hyperlink{cite.3gpp38901}{TR~38.901}, Sec.~7.6.13, is set to zero for every link. The three sub-cluster delays of Eq.~(7.5-26) are also omitted, so all rays in a cluster share this source distance. With $\mathrm{dist}_{k,b,g,s}^{(c,\ell)}$ the distance from source $(c,\ell)$ to element $(g,s)$, the ray response is
\begin{equation}
 \left[\mathbf{a}_{k,b,g}^{(c,\ell)}\right]_{s}
 =
 \frac{d^{1}_{k,b,c}}{\mathrm{dist}_{k,b,g,s}^{(c,\ell)}}
 A^{E,(c,\ell)}_{k,b,g,s}
 \sqrt{\chi_{k,b,g,s}^{(c,\ell)}}\,
 e^{-\jmath 2\pi\,\mathrm{dist}_{k,b,g,s}^{(c,\ell)}/\lambda}.
\end{equation} The LoS and NLoS element-wise spreading factors form the adopted near-field amplitude model, beyond the constant-amplitude Step~11 coefficient. With $\chi_{k,b,g,s}^{(c,\ell)}\in(0,1]$ the SNS factor and $\alpha_{k,b,c,\ell}\in\mathbb{C}$ the gain of the UE-to-scatterer segment, the NLoS blocks are
\begin{equation}
 \mathbf{h}_{k,b,g}^{\mathrm{NLoS},c}
 =
 \sum_{\ell=1}^{L_{k,b,c}}
 \alpha_{k,b,c,\ell}\,
 \mathbf{a}_{k,b,g}^{(c,\ell)},
 \qquad
 \mathbf{h}_{k,b,g}^{\mathrm{NLoS}}
 =
 \sum_{c\in\mathcal{C}_{k,b}}
 \mathbf{h}_{k,b,g}^{\mathrm{NLoS},c}.
 \label{eq:nlos} \end{equation} By \hyperlink{assumption:A5}{(A5)}, $\mathbb{E}[\alpha_{k,b,c,\ell}]=0$ and
$\mathbb{E}[\alpha_{k,b,c,\ell}
\alpha_{k,b,c',\ell'}^{*}]
=\sigma_{k,b,c,\ell}^{2}\,
\mathbf{1}\{(c,\ell)=(c',\ell')\}$.
The factors $\chi$ use the physical-blocker SNS construction of \hyperlink{cite.3gpp38901}{TR~38.901} Rel-19, Sec.~7.6.14.1.2, with the diffraction approximation specified in \hyperref[sec:setup]{Section~\ref*{sec:setup}}, or its stochastic model in Sec.~7.6.14.1.3. The positive $\chi$ factors do not impose hard zeros on individual rays. Stacking the blocks, the true channel is
\begin{equation}
 \mathbf{h}_{k,b}
 =
 \sqrt{\varrho_{k,b}}\left(
 \sqrt{\frac{1}{1+\mathcal{K}_{k,b}}}\,
 \mathbf{h}_{k,b}^{\mathrm{NLoS}}
 +
 \sqrt{\frac{\mathcal{K}_{k,b}}{1+\mathcal{K}_{k,b}}}\,
 \mathbf{h}_{k,b}^{\mathrm{LoS}}
 \right),
\end{equation} with link gain $\varrho_{k,b}$ (\hyperlink{cite.3gpp38901}{TR~38.901}, Table~7.4.1-1) and Ricean factor $\mathcal{K}_{k,b}$. With block energy $\beta_{k,b,g}\triangleq\mathbb{E}\|\mathbf{h}_{k,b,g}\|_{2}^{2}$, where the expectation is over small-scale gains conditional on the large-scale state, the VR of UE $k$ at BS $b$ is
\begin{equation}
 \begin{split}
 \mathcal{V}_{k,b}
 \triangleq
 \Big\{
 g\in\mathcal{G}:\
 &\beta_{k,b,g}
 \geq
 10^{-\eta/10}\max\nolimits_{g'\in\mathcal{G}}\beta_{k,b,g'},\\
 &\beta_{k,b,g}P_{b}^{\max}/\sigma_{k}^{2}
 \geq
 10^{\gamma_{\min}/10}
 \Big\}.
 \end{split}
 \label{eq:vr_set} \end{equation} The nominal INR gate is $\gamma_{\min}=-10$~dB; it tests potential interference at the maximum BS power, not measured instantaneous interference. The relative threshold $\eta$ is the design knob. A VR may be empty. The gating operator is
\begin{equation}
 \mathbf{D}_{k,b}
 \triangleq
 \operatorname{blkdiag}\!\left(
 \mathbf{1}\{1\in\mathcal{V}_{k,b}\}\mathbf{I}_{N_{\mathrm{s}}},
 \ldots,
 \mathbf{1}\{G\in\mathcal{V}_{k,b}\}\mathbf{I}_{N_{\mathrm{s}}}
 \right),
\end{equation} and the design channel is
\begin{equation}
 \tilde{\mathbf{h}}_{k,b}
 \triangleq
 \mathbf{D}_{k,b}\mathbf{h}_{k,b}.
 \label{eq:gated_channel} \end{equation} The residual $\mathbf{r}_{k,b}=\mathbf{h}_{k,b}-\tilde{\mathbf{h}}_{k,b}$ is bounded in \hyperref[sec:robust]{Section~\ref*{sec:robust}}.

\subsection{Multi-Cell VR Conflict Graph and Victim Sets}
\label{sec:conflict_graph}

At each BS $b$, define the VR conflict graph
\begin{equation}
 \mathcal{G}_{b}^{\mathrm{c}}
 \triangleq
 \left(\mathcal{K},\mathcal{E}_{b}\right),
\end{equation} where, for $k\neq j$,
\begin{equation}
 \{k,j\}\in\mathcal{E}_{b}
 \quad\Longleftrightarrow\quad
 \mathcal{V}_{k,b}\cap\mathcal{V}_{j,b}\neq\emptyset.
 \label{eq:conflict_edge} \end{equation} Edges may join UEs of any cells. The victim set of BS $b$ is
\begin{equation}
 \mathcal{K}_{b}^{\mathrm{vic}}
 \triangleq
 \left\{
 k\in\mathcal{K}:
 a_{k}\neq b,\;
 \mathcal{V}_{k,b}\neq\emptyset
 \right\},
 \label{eq:victim_set} \end{equation} and its effective victim set is
\begin{equation}
 \mathcal{K}_{b}^{\mathrm{vic,eff}}
 \triangleq
 \left\{
 k\in\mathcal{K}_{b}^{\mathrm{vic}}:
 \exists\, j\in\mathcal{K}_{b}
 \ \mathrm{such\ that}\
 \{k,j\}\in\mathcal{E}_{b}
 \right\}.
 \label{eq:effective_victim_set} \end{equation}

\subsection{Beamformer Support, Received Signal, and SINR}
\label{sec:sinr}

Let $\mathbf{w}_{k,b}\in\mathbb{C}^{N}$, with blocks $\mathbf{w}_{k,b,g}$, be the beamformer of BS $b$ for UE $k$, with subarray support
\begin{equation}
 \operatorname{supp}_{\mathrm{sub}}(\mathbf{w}_{k,b})
 \triangleq
 \left\{
 g\in\mathcal{G}:
 \mathbf{w}_{k,b,g}\neq\mathbf{0}
 \right\}.
\end{equation} Under CB and the VR support constraint,
\begin{equation}
 \mathbf{w}_{k,b}=\mathbf{0}
 \quad\text{if } b\neq a_{k},
 \qquad
 \operatorname{supp}_{\mathrm{sub}}(\mathbf{w}_{k,a_{k}})
 \subseteq
 \mathcal{V}_{k,a_{k}}.
 \label{eq:support} \end{equation} The per-BS power constraint is
\begin{equation}
 \sum_{k\in\mathcal{K}_{b}}
 \left\|\mathbf{w}_{k,b}\right\|_{2}^{2}
 \leq P_{b}^{\max},
 \qquad b\in\mathcal{B}.
\end{equation} With data symbols $s_{k}$, the design-model received signal at UE $k$ is
\begin{align}
 y_{k}
 ={}&
 \tilde{\mathbf{h}}_{k,a_{k}}^{\mathrm{H}}
 \mathbf{w}_{k,a_{k}}s_{k}
 +
 \sum_{\substack{j\in\mathcal{K}_{a_{k}}\\ j\neq k}}
 \tilde{\mathbf{h}}_{k,a_{k}}^{\mathrm{H}}
 \mathbf{w}_{j,a_{k}}s_{j}
 \nonumber\\
 &+
 \sum_{b\in\mathcal{B}\setminus\{a_{k}\}}
 \sum_{j\in\mathcal{K}_{b}}
 \tilde{\mathbf{h}}_{k,b}^{\mathrm{H}}
 \mathbf{w}_{j,b}s_{j}
 +
 n_{k},
 \label{eq:received} \end{align} with intra- and inter-cell interference powers
\begin{align}
 I_{k}^{\mathrm{intra}}
 &\triangleq
 \sum_{\substack{j\in\mathcal{K}_{a_{k}}\\ j\neq k}}
 \left|
 \tilde{\mathbf{h}}_{k,a_{k}}^{\mathrm{H}}
 \mathbf{w}_{j,a_{k}}
 \right|^{2},
 \\
 I_{k}^{\mathrm{inter}}
 &\triangleq
 \sum_{b\in\mathcal{B}\setminus\{a_{k}\}}
 \sum_{j\in\mathcal{K}_{b}}
 \left|
 \tilde{\mathbf{h}}_{k,b}^{\mathrm{H}}
 \mathbf{w}_{j,b}
 \right|^{2}.
\end{align} The design SINR and rate are
\begin{equation}
 \gamma_{k}
 =
 \frac{
 \left|
 \tilde{\mathbf{h}}_{k,a_{k}}^{\mathrm{H}}
 \mathbf{w}_{k,a_{k}}
 \right|^{2}
 }{
 I_{k}^{\mathrm{intra}}+I_{k}^{\mathrm{inter}}+\sigma_{k}^{2}
 },
 \qquad
 R_{k}=\log_{2}(1+\gamma_{k}).
 \label{eq:sinr} \end{equation} They are denoted $\hat{\Gamma}_{k}$ and $\hat{R}_{k}$ in \hyperref[sec:robust]{Section~\ref*{sec:robust}}. On the true channel,
\begin{equation}
 y_{k}^{\mathrm{true}}
 =
 \mathbf{h}_{k,a_{k}}^{\mathrm{H}}
 \mathbf{w}_{k,a_{k}}s_{k}
 +
 \sum_{\substack{b\in\mathcal{B},\, j\in\mathcal{K}_{b}\\
 (b,j)\neq(a_{k},k)}}
 \mathbf{h}_{k,b}^{\mathrm{H}}\mathbf{w}_{j,b}s_{j}
 +
 n_{k},
 \label{eq:received_true} \end{equation} and the true SINR and rate, reported in \hyperref[sec:results]{Section~\ref*{sec:results}}, are
\begin{equation}
 \Gamma_{k}
 =
 \frac{
 \left|
 \mathbf{h}_{k,a_{k}}^{\mathrm{H}}
 \mathbf{w}_{k,a_{k}}
 \right|^{2}
 }{
 \sum_{\substack{b\in\mathcal{B},\, j\in\mathcal{K}_{b}\\
 (b,j)\neq(a_{k},k)}}
 \left|
 \mathbf{h}_{k,b}^{\mathrm{H}}\mathbf{w}_{j,b}
 \right|^{2}
 +\sigma_{k}^{2}
 },
 \label{eq:sinr_true} \end{equation}
and $R_{k}^{\mathrm{true}}=\log_{2}(1+\Gamma_{k})$.

\subsection{BS-to-CU CSI Upload Load}
\label{sec:csi_exchange}

Per BS, the CSI uploaded to the CU, in complex values, is: (a) the full channel of every UE,
\begin{equation}
 \mathcal{L}_{b}^{\mathrm{full}}
 =
 KN
 =
 KGN_{\mathrm{s}}
\end{equation}
(b) the full channel of each UE with a nonempty VR,
\begin{equation}
 \mathcal{L}_{b}^{\mathrm{UE}}
 =
 \left|
 \left\{
 k\in\mathcal{K}:\mathcal{V}_{k,b}\neq\emptyset
 \right\}
 \right| N
\end{equation}
(c) the nonzero blocks of the design channels,
\begin{equation}
 \mathcal{L}_{b}^{\mathrm{VR}}
 =
 \sum_{k\in\mathcal{K}}
 \left|\mathcal{V}_{k,b}\right|N_{\mathrm{s}}
 \label{eq:csi_vr}
\end{equation}
plus $KG$ mask bits. Since $|\mathcal{V}_{k,b}|\leq G$ and at most $K$ VRs are nonempty,
\begin{equation}
 \mathcal{L}_{b}^{\mathrm{VR}}
 \leq
 \mathcal{L}_{b}^{\mathrm{UE}}
 \leq
 \mathcal{L}_{b}^{\mathrm{full}}.
\end{equation} Load ratios count complex channel entries only, excluding mask bits, quantization, pilots, and transport framing.

\subsection{Structural Lemmas}
\label{sec:lemmas}

Both lemmas concern the design channel. Let
\begin{equation}
 \mathbf{R}_{k,b}
 \triangleq
 \mathbb{E}\!\left[
 \left(
 \tilde{\mathbf{h}}_{k,b}
 -
 \mathbb{E}[\tilde{\mathbf{h}}_{k,b}]
 \right)
 \left(
 \tilde{\mathbf{h}}_{k,b}
 -
 \mathbb{E}[\tilde{\mathbf{h}}_{k,b}]
 \right)^{\mathrm{H}}
 \right].
\end{equation} 

\begin{lemma}[Rank of the VR-gated channel covariance]
\label{lem:rank}
Fix $(k,b)$. Then
$\mathbf{R}_{k,b}$ has zero rows and zero columns for every subarray
$g\notin\mathcal{V}_{k,b}$, and
\begin{equation}
 \operatorname{rank}(\mathbf{R}_{k,b})
 \leq
 \min\!\left\{
 \left|\mathcal{V}_{k,b}\right|N_{\mathrm{s}},
 \;
 \sum_{c\in\mathcal{C}_{k,b}} L_{k,b,c}
 \right\}.
 \label{eq:rank_bound} \end{equation} \end{lemma}

\noindent\textit{Proof:} see \hyperlink{proof:A}{Appendix~\ref*{app:proofs}}.

Hence the random part of the gated channel lives in a subspace whose dimension is limited by both the visible entries and the number of rays, so second-order CSI for link $(k,b)$ never requires coordinates outside $\mathcal{V}_{k,b}$.

Define the same-BS intersection
\begin{equation}
 \mathcal{I}_{k,j,b}
 \triangleq
 \mathcal{V}_{k,b}\cap\mathcal{V}_{j,b}.
\end{equation}
Define $\tilde{\mathbf{h}}_{k,b}^{\mathrm{NLoS}}\triangleq\sqrt{\varrho_{k,b}/(1+\mathcal{K}_{k,b})}\,\mathbf{D}_{k,b}\mathbf{h}_{k,b}^{\mathrm{NLoS}}$, with block $g$ denoted $\tilde{\mathbf{h}}_{k,b,g}^{\mathrm{NLoS}}$.

\begin{lemma}[Interference bound scaled by subarray intersection]
\label{lem:interference}
Fix $(k,j,b)$ and suppose $\mathbf{w}_{j,b}$ satisfies the support
constraint \eqref{eq:support}. Define
\begin{equation}
 A_{k,b,g}^{\mathrm{LoS}}
 \triangleq
 \max_{1\leq s\leq N_{\mathrm{s}}}
 \left|
 \left[
 \sqrt{\varrho_{k,b}}\sqrt{\tfrac{\mathcal{K}_{k,b}}
 {1+\mathcal{K}_{k,b}}}\,
 \mathbf{h}_{k,b,g}^{\mathrm{LoS}}
 \right]_{s}
 \right|.
\end{equation}
Then
\begin{align}
 &\left|
 \tilde{\mathbf{h}}_{k,b}^{\mathrm{H}}\mathbf{w}_{j,b}
 \right|^{2}
 \nonumber\\
 &\quad\leq
 \left\|\mathbf{w}_{j,b}\right\|_{2}^{2}
 \sum_{g\in\mathcal{I}_{k,j,b}}
 \left\|
 \tilde{\mathbf{h}}_{k,b,g}
 \right\|_{2}^{2}
 \label{eq:interference_bound_1}\\
 &\quad\leq
 2\left\|\mathbf{w}_{j,b}\right\|_{2}^{2}
 \sum_{g\in\mathcal{I}_{k,j,b}}
 \left[
 N_{\mathrm{s}}
 \left(A_{k,b,g}^{\mathrm{LoS}}\right)^{2}
 +
 \left\|
 \tilde{\mathbf{h}}_{k,b,g}^{\mathrm{NLoS}}
 \right\|_{2}^{2}
 \right]
 \label{eq:interference_bound_2}\\
 &\quad\leq
 2\left|\mathcal{I}_{k,j,b}\right|
 \left\|\mathbf{w}_{j,b}\right\|_{2}^{2}
 \max_{g\in\mathcal{I}_{k,j,b}}
 \left[
 N_{\mathrm{s}}
 \left(A_{k,b,g}^{\mathrm{LoS}}\right)^{2}
 +
 \left\|
 \tilde{\mathbf{h}}_{k,b,g}^{\mathrm{NLoS}}
 \right\|_{2}^{2}
 \right],
 \label{eq:interference_bound_3}
\end{align}
where the maximum in \eqref{eq:interference_bound_3} is read as
zero when $\mathcal{I}_{k,j,b}=\emptyset$. In particular, if
$\mathcal{I}_{k,j,b}=\emptyset$, then
$\tilde{\mathbf{h}}_{k,b}^{\mathrm{H}}\mathbf{w}_{j,b}=0$.
\end{lemma}

\noindent\textit{Proof:} see \hyperlink{proof:B}{Appendix~\ref*{app:proofs}}.

\subsection{Problem Formulation}
\label{sec:problem}

The joint association and CB design is
\begin{equation}
\begin{aligned}
\max_{\{a_{k}\},\{\mathbf{w}_{k,b}\}}\quad &
\textstyle\sum_{k\in\mathcal{K}}\omega_{k}\log_{2}(1+\gamma_{k})\\
\mathrm{s.t.}\quad &
a_{k}\in\{b\in\mathcal{B}:\mathcal{V}_{k,b}\neq\emptyset\}, && \forall k,\\
& \mathbf{w}_{k,b}=\mathbf{0}, && \forall k,\ \forall b\neq a_{k},\\
& \mathbf{w}_{k,b,g}=\mathbf{0}, && \forall k,b,\ \forall g\notin\mathcal{V}_{k,b},\\
& \textstyle\sum_{k\in\mathcal{K}_{b}}\|\mathbf{w}_{k,b}\|_{2}^{2}\leq P_{b}^{\max}, && \forall b.
\end{aligned}
\label{prob:main} \end{equation} Problem~\eqref{prob:main} is mixed-integer and nonconvex. \hyperref[alg:main]{Algorithm~\ref*{alg:main}} uses a gain-based association heuristic, then optimizes the beams for that fixed association. \hyperref[sec:robust]{Section~\ref*{sec:robust}} bounds the design-to-true gap.
\section{Proposed Algorithm}
\label{sec:algorithm}

Let $\nu_{k}>0$ be the MMSE weight of UE $k$ (distinct from the rate weight $\omega_{k}$), $u_{k}\in\mathbb{C}$ its MMSE receiver, and $\mu_{b}\geq 0$ the power multiplier of BS $b$. The served support of BS $b$ and the conflict neighborhood
of UE $i$ at BS $b$ are
\begin{equation}
 \mathcal{U}_{b}
 \triangleq
 \bigcup_{i\in\mathcal{K}_{b}}\mathcal{V}_{i,b},
 \qquad
 \mathcal{N}_{b}(i)
 \triangleq
 \left\{
 j\in\mathcal{K}:
 j=i\ \text{or}\ \{i,j\}\in\mathcal{E}_{b}
 \right\}.
\end{equation}

\emph{Timescales.}\label{sec:timescales} Association uses only block gains and VR masks; instantaneous CSI drives the subsequent masked WMMSE updates. At fixed channels, power gate, and association, increasing $\eta$ expands the VRs and cannot reduce the counted loads. Reassociation can change this ordering for conflict-aware load, and true-channel rates need not increase with $\eta$.

\subsection{Masked Multi-Cell WMMSE}
\label{sec:wmmse}

Fix the association $\{a_{k}\}$. Write $\mathbf{w}_{i}\triangleq\mathbf{w}_{i,a_{i}}$ and let
$\mathcal{S}_{i}\triangleq
\{\mathbf{w}\in\mathbb{C}^{N}:
\mathbf{w}_{g}=\mathbf{0}\ \forall g\notin\mathcal{V}_{i,a_{i}}\}$
be the support subspace of stream $i$. The feasible beamformer set
is the product over $b\in\mathcal{B}$ of
$\{\{\mathbf{w}_{i}\}_{i\in\mathcal{K}_{b}}:
\mathbf{w}_{i}\in\mathcal{S}_{i},
\sum_{i\in\mathcal{K}_{b}}\|\mathbf{w}_{i}\|_{2}^{2}
\leq P_{b}^{\max}\}$.

\emph{Initialization.} Each stream is initialized by restricted MRT on its VR
support with equal power per BS
(\hyperref[alg:main]{Algorithm~\ref*{alg:main}}). A zero-norm support gives
$\mathbf{w}_{i}=\mathbf{0}$.

\subsubsection{WMMSE reformulation}
For UE $k$, let the scalar estimate be
$\hat{s}_{k}=u_{k}^{*}y_{k}$, with $y_{k}$ as in
\eqref{eq:received}. Its mean-square error is
\begin{align}
 e_{k}
 &\triangleq
 \mathbb{E}\!\left[\left|u_{k}^{*}y_{k}-s_{k}\right|^{2}\right]
 \nonumber\\
 &=
 |u_{k}|^{2}Q_{k}
 -
 2\operatorname{Re}\!\left\{
 u_{k}^{*}
 \tilde{\mathbf{h}}_{k,a_{k}}^{\mathrm{H}}\mathbf{w}_{k}
 \right\}
 +1,
 \label{eq:mse}
\end{align}
where
$Q_{k}\triangleq
\sum_{j\in\mathcal{K}}
|\tilde{\mathbf{h}}_{k,a_{j}}^{\mathrm{H}}\mathbf{w}_{j}|^{2}
+\sigma_{k}^{2}$
is the total received power at UE $k$. Following \cite{shi2011}, fixed-association Problem~\eqref{prob:main} is equivalent to minimizing
\begin{equation}
 \mathcal{F}(\{\mathbf{w}_{i}\},\{u_{k}\},\{\nu_{k}\})
 \triangleq
 \sum_{k\in\mathcal{K}}
 \omega_{k}
 \left(
 \nu_{k}e_{k}-\ln\nu_{k}
 \right)
 \label{eq:wmmse_aux}
\end{equation}
over $\mathbf{w}_{i}\in\mathcal{S}_{i}$, $u_{k}$ and $\nu_{k}>0$, subject to the per-BS power constraints. At the minimizing $(u_{k},\nu_{k})$ the auxiliary objective equals $\sum_k\omega_k-(\ln2)\sum_k\omega_k R_k$; minimizing it maximizes the design weighted sum rate. This follows from $e_k^{\min}=1/(1+\gamma_k)$.

\subsubsection{Receiver and weight updates}
Minimizing \eqref{eq:mse} over $u_{k}$ gives
\begin{equation}
 u_{k}
 =
 \frac{
 \tilde{\mathbf{h}}_{k,a_{k}}^{\mathrm{H}}\mathbf{w}_{k}
 }{
 \displaystyle
 \sum_{j\in\mathcal{K}}
 \left|
 \tilde{\mathbf{h}}_{k,a_{j}}^{\mathrm{H}}\mathbf{w}_{j}
 \right|^{2}
 +\sigma_{k}^{2}
 }.
 \label{eq:u_update}
\end{equation}
Substitution into \eqref{eq:mse} gives
$e_{k}^{\min}=1-|\tilde{\mathbf{h}}_{k,a_{k}}^{\mathrm{H}}
\mathbf{w}_{k}|^{2}/Q_{k}=1/(1+\gamma_{k})$, with $\gamma_{k}$ as
in \eqref{eq:sinr}. Minimizing
$\omega_{k}(\nu_{k}e_{k}-\ln\nu_{k})$ over $\nu_{k}>0$ then gives
\begin{equation}
 \nu_{k}=\frac{1}{e_{k}},
 \label{eq:nu_update}
\end{equation}
evaluated at the updated $u_{k}$, hence
$\nu_{k}=1+\gamma_{k}$. The implementation guards the
division as $\nu_{k}=1/\max(e_{k},10^{-12})$.

\subsubsection{Beamformer update}
Fix $\{u_{k}\}$ and $\{\nu_{k}\}$. In \eqref{eq:wmmse_aux}, the
terms that depend on $\mathbf{w}_{i}$, where $b=a_{i}$, are
\begin{equation}
 \mathbf{w}_{i}^{\mathrm{H}}
 \mathbf{A}_{i,b}^{\mathrm{full}}
 \mathbf{w}_{i}
 -
 2\operatorname{Re}\!\left\{
 \left(
 \omega_{i}\nu_{i}u_{i}\tilde{\mathbf{h}}_{i,b}
 \right)^{\mathrm{H}}
 \mathbf{w}_{i}
 \right\},
 \label{eq:beam_terms}
\end{equation}
where
$\mathbf{A}_{i,b}^{\mathrm{full}}
\triangleq
\sum_{j\in\mathcal{K}}
\omega_{j}\nu_{j}|u_{j}|^{2}
\tilde{\mathbf{h}}_{j,b}
\tilde{\mathbf{h}}_{j,b}^{\mathrm{H}}$.
The subproblem separates by BS and, for fixed $\mu_{b}$, across streams; each stream is minimized exactly over $\mathcal{S}_{i}$.

Let $\mathcal{V}_{i}\triangleq\mathcal{V}_{i,b}$,
$m_{i}=|\mathcal{V}_{i}|N_{\mathrm{s}}$, and, for any vector
$\mathbf{z}$ supported on the blocks of $\mathcal{V}_{i}$, let
$\mathbf{z}_{\mathcal{V}_{i}}\in\mathbb{C}^{m_{i}}$ be its subvector
on the coordinates of $\mathcal{V}_{i}$. Restricting
$\mathbf{A}_{i,b}^{\mathrm{full}}$ to these coordinates gives the
principal submatrix
\begin{equation}
 \mathbf{A}_{i,b}
 \triangleq
 \sum_{j\in\mathcal{N}_{b}(i)}
 \omega_{j}\nu_{j}|u_{j}|^{2}\,
 \tilde{\mathbf{h}}_{j,b,\mathcal{V}_{i}}
 \tilde{\mathbf{h}}_{j,b,\mathcal{V}_{i}}^{\mathrm{H}}.
 \label{eq:principal_submatrix}
 \end{equation}
The sum reduces to $\mathcal{N}_{b}(i)$ because, by \eqref{eq:gated_channel}, $\tilde{\mathbf{h}}_{j,b}$ vanishes on $\mathcal{V}_{i}$ for $j\notin\mathcal{N}_{b}(i)$. In exact arithmetic, the restricted-subproblem minimizer for $\mu_{b}>0$ is
\begin{equation}
 \mathbf{w}_{i,b,\mathcal{V}_{i}}
 =
 \left(
 \mathbf{A}_{i,b}+\mu_{b}\mathbf{I}_{m_{i}}
 \right)^{-1}
 \left(
 \omega_{i}\nu_{i}u_{i}\,
 \tilde{\mathbf{h}}_{i,b,\mathcal{V}_{i}}
 \right),
 \label{eq:beam_update}
\end{equation}
embedded into $\mathbb{C}^{N}$ with zeros elsewhere. For $\mu_{b}=0$, \eqref{eq:beam_update} is read with the Moore--Penrose pseudoinverse $\mathbf{A}_{i,b}^{+}$ in place of the inverse, the minimum-norm minimizer when $\mathbf{A}_{i,b}$ is singular.

The implementation uses one eigendecomposition $\mathbf A_{i,b}=\mathbf U_i\boldsymbol\Lambda_i\mathbf U_i^{\mathrm H}$ per stream, with $\mathbf d_i=\mathbf U_i^{\mathrm H}(\omega_i\nu_i u_i\tilde{\mathbf h}_{i,b,\mathcal V_i})$. The right-hand side lies in the range of $\mathbf A_{i,b}$ because the sum includes $j=i$. Numerically, components with $\lambda_{i,r}\leq10^{-10}\max_{r'}\lambda_{i,r'}$ are discarded, including at $\mu_b=0$.

The multiplier $\mu_{b}$ is selected by bisection on
$P_b(\mu)=\sum_{i\in\mathcal K_b}\sum_r |d_{i,r}|^2/(\lambda_{i,r}+\mu)^2$, with the stated pseudoinverse convention at zero and discarded components omitted. It is $\mu_{b}=0$ if
$P_{b}(0)\leq P_{b}^{\max}$. Otherwise the upper bracket
is the closed form $\mu_{\mathrm{hi}}=\sqrt{\sum_{i\in\mathcal{K}_{b}} \|\mathbf{d}_{i}\|_{2}^{2}/P_{b}^{\max}}$.
This bracket satisfies
$P_{b}(\mu_{\mathrm{hi}})\leq P_{b}^{\max}$ term by term.
The bisection runs to a relative tolerance of $10^{-9}$
in at most $60$ steps. It returns $\mu_{\mathrm{hi}}$,
so the power constraint is met to numerical tolerance.

Masking the unrestricted solution instead is not an exact restricted solve in general. Block diagonality of $\mathbf{A}_{i,b}^{\mathrm{full}}$ with respect to $\mathcal{V}_{i}$ and its complement is a sufficient condition for equivalence, but is not necessary for a particular right-hand side.

\begin{proposition}[Monotonicity of the safeguarded iteration] \normalfont
\label{prop:wmmse_convergence}
Fix a feasible association. Consider the iteration of
this section as implemented. It uses the updates
\eqref{eq:u_update}, \eqref{eq:nu_update} and
\eqref{eq:beam_update} with the bisection above. A
safeguard then discards the new beamformers and stops if
the design weighted sum rate has decreased. Then the design weighted sum rate sequence is
non-decreasing and bounded above, and therefore converges.
This is convergence of objective values; it does not establish stationary-point convergence of the numerically truncated, safeguarded implementation. The iteration stops when the design weighted sum rate increase
is below $10^{-4}$ or after $60$ iterations.
\end{proposition}

\noindent\textit{Proof:} see \hyperlink{proof:C}{Appendix~\ref*{app:proofs}}.

When $\mathcal N_b(i)=\{i\}$, the update reduces to restricted MRT, with power set by $\mu_b$. Otherwise, each neighboring UE contributes a rank-one penalty on its shared blocks, including effective inter-cell victims.

\subsection{CSI Dependence of the Algorithm}
\label{sec:csi_dependence}

\begin{proposition}[Channel blocks used by the updates] \normalfont
\label{prop:csi_dependence}
Fix the association. For each BS $b$, every
fast-timescale update of \hyperref[alg:main]{Algorithm~\ref*{alg:main}} depends
on the channels indexed by $b$ only through the blocks
\begin{equation}
 \left\{
 \tilde{\mathbf{h}}_{j,b,g}:
 j\in\mathcal{K},\;
 g\in\mathcal{V}_{j,b}\cap\mathcal{U}_{b}
 \right\}.
  \label{eq:conflict_blocks}
\end{equation}
The updates are $\{u_{k}\}$, $\{\nu_{k}\}$,
$\{\mathbf{w}_{i}\}$ and $\{\mu_{b}\}$.
\end{proposition}

\noindent\textit{Proof:} see \hyperlink{proof:D}{Appendix~\ref*{app:proofs}}.

\begin{corollary}[Conflict-aware upload load]
\label{cor:conflict_load}
For fixed association, define the conflict-aware load of BS $b$ by
\begin{equation}
 \mathcal{L}_{b}^{\mathrm{conf}}
 \triangleq
 \sum_{j\in\mathcal{K}}
 \left|
 \mathcal{V}_{j,b}\cap\mathcal{U}_{b}
 \right|N_{\mathrm{s}}.
 \label{eq:csi_conf}
\end{equation}
Then
$\mathcal{L}_{b}^{\mathrm{conf}}\leq\mathcal{L}_{b}^{\mathrm{VR}}$,
with $\mathcal{L}_{b}^{\mathrm{VR}}$ as in
\eqref{eq:csi_vr}. Equality holds if and only if
$\mathcal{V}_{j,b}\subseteq\mathcal{U}_{b}$ for every
$j\in\mathcal{K}$.
\end{corollary}

\noindent\textit{Proof:} see \hyperlink{proof:E}{Appendix~\ref*{app:proofs}}.

A strict gap
$\mathcal{L}_{b}^{\mathrm{conf}}<\mathcal{L}_{b}^{\mathrm{VR}}$
requires a UE visible to BS $b$ on blocks outside
$\mathcal{U}_{b}$. Since $\mathcal{U}_{b}$ grows with
$|\mathcal{K}_{b}|$, the gap is load dependent
(\hyperref[sec:expected_loads]{Section~\ref*{sec:expected_loads}}).

The load \eqref{eq:csi_conf} depends on the association through $\mathcal{U}_{b}$, so it applies on the fast timescale. The slow timescale uses no instantaneous CSI. Per BS it
requires only the visible gains $\{\beta_{k,b,g}\}$,
$KG$ real values, and the UE VR masks, $KG$ bits.

\subsection{Association}
\label{sec:association}

Each UE is initialized at the feasible BS with the largest total
visible gain:
\begin{equation}
 a_{k}
 =
 \underset{
 b\in\{b':\mathcal{V}_{k,b'}\neq\emptyset\}
 }{\arg\max}
 \;
 \sum_{g\in\mathcal{V}_{k,b}}
 \beta_{k,b,g},
 \qquad k\in\mathcal{K}.
 \label{eq:assoc_init}
\end{equation}
Ties in \eqref{eq:assoc_init} favor the smallest BS index. Each included UE must have a nonempty VR at some BS; UEs with all-empty VRs are excluded from every scheme (\hyperref[sec:setup]{Section~\ref*{sec:setup}}).

\subsection{Complexity}
\label{sec:complexity}

Let $m_{i}=|\mathcal{V}_{i,a_{i}}|N_{\mathrm{s}}$ be the
support size of stream $i$. Let $T_{\mu}$ be the number
of bisection trials for a multiplier $\mu_{b}$ in one
fast iteration. It is at most $60$.

The receiver and weight updates cost
$\mathcal{O}\big(\sum_{k}\sum_{j}
|\mathcal{V}_{k,a_{j}}\cap\mathcal{V}_{j,a_{j}}|
N_{\mathrm{s}}\big)$ in the masked iteration, against
$\mathcal{O}(K^{2}N)$ for full-aperture WMMSE. The beamformer updates use one eigendecomposition per
stream. Every bisection trial is then linear in the
support sizes. The beamformer cost at BS $b$ is
\begin{equation}
 \mathcal{O}\!\left(
 \sum_{i\in\mathcal{K}_{b}}
 \left(
 |\mathcal{N}_{b}(i)|m_{i}^{2}+m_{i}^{3}
 \right)
 +
 T_{\mu}
 \sum_{i\in\mathcal{K}_{b}}m_{i}
 \right).
 \label{eq:complexity_masked}
\end{equation}
For full-aperture WMMSE, the same computation with $m_{i}=N$ and sums over all $K$ receivers gives $\mathcal{O}\big(|\mathcal{K}_{b}|KN^{2} +|\mathcal{K}_{b}|N^{3} +T_{\mu}|\mathcal{K}_{b}|N\big)$ at BS $b$. The masked cost replaces $N$ and $K$ by the
support and neighborhood sizes.

\begin{algorithm}[t]
\caption{Two-timescale masked coordinated beamforming}
\label{alg:main}
\begin{algorithmic}[1]
\REQUIRE VR sets $\{\mathcal{V}_{k,b}\}$, block gains
$\{\beta_{k,b,g}\}$, instantaneous channels
$\{\tilde{\mathbf{h}}_{k,b}\}$, weights $\{\omega_{k}\}$, powers
$\{P_{b}^{\max}\}$
\STATE \textbf{Slow timescale: association}
\STATE Initialize $a_{k}$ by \eqref{eq:assoc_init}; resolve ties
by the smallest BS index
\STATE \textbf{Fast timescale: fixed-association WMMSE}
\STATE Initialize each $\mathbf{w}_{i}$, $i\in\mathcal{K}_{b}$, as restricted MRT with equal power on $\mathcal{V}_{i,b}$ (\hyperref[sec:wmmse]{Section~\ref*{sec:wmmse}}), or $\mathbf{0}$ if the norm is zero; set $\mathbf{w}_{k,b}=\mathbf{0}$ for $b\neq a_{k}$
\STATE Set $R^{(0)}$ to the design weighted sum rate at the initial
beamformers
\FOR{$t=1,2,\ldots,60$}
\STATE Update $\{u_{k}\}$ by \eqref{eq:u_update} and
$\{\nu_{k}\}$ by \eqref{eq:nu_update}
\FOR{each $b\in\mathcal{B}$}
\STATE Form $\{\mathbf{A}_{i,b}\}_{i\in\mathcal{K}_{b}}$ by \eqref{eq:principal_submatrix}; compute one eigendecomposition of each $\mathbf{A}_{i,b}$ and the rotated right-hand sides $\mathbf{d}_{i}$, zeroing components with $\lambda_{i,r}\leq 10^{-10}\max_{r'}\lambda_{i,r'}$
\STATE Select $\mu_{b}$ by the bisection of \hyperref[sec:wmmse]{Section~\ref*{sec:wmmse}}
\STATE Update $\{\mathbf{w}_{i}\}_{i\in\mathcal{K}_{b}}$ by
\eqref{eq:beam_update} with this $\mu_{b}$
\ENDFOR
\STATE Compute the design weighted sum rate $R^{(t)}$ from
\eqref{eq:sinr}
\IF{$R^{(t)}<R^{(t-1)}$}
\STATE Restore the previous beamformers and \textbf{break}
\ENDIF
\IF{$R^{(t)}-R^{(t-1)}<10^{-4}$}
\STATE \textbf{break}
\ENDIF
\ENDFOR
\RETURN $\{a_{k}\}$ and $\{\mathbf{w}_{k,b}\}$
\end{algorithmic}
\end{algorithm}
\section{Structural Analysis}
\label{sec:analysis}

We state when \hyperref[alg:main]{Algorithm~\ref*{alg:main}} reduces to per-cell beamforming and give its expected CSI loads, using
\begin{equation}
 q_{k,b,g}
 \triangleq
 \Pr\!\left(g\in\mathcal{V}_{k,b}\right),
 \qquad
 \zeta_{b,g}
 \triangleq
 1-\prod_{i\in\mathcal{K}_{b}}\left(1-q_{i,b,g}\right).
\end{equation}

\subsection{Exactness of Per-Cell Decoupling}
\label{sec:decoupling}

The per-cell masked WMMSE baseline uses the same updates
as \hyperref[alg:main]{Algorithm~\ref*{alg:main}}, including the safeguard. Its
beamformer matrix for stream $i$ at BS $b$ is
\begin{equation}
 \mathbf{A}_{i,b}^{\mathrm{pc}}
 \triangleq
 \sum_{j\in\mathcal{N}_{b}(i)\cap\mathcal{K}_{b}}
 \omega_{j}\nu_{j}|u_{j}|^{2}\,
 \tilde{\mathbf{h}}_{j,b,\mathcal{V}_{i}}
 \tilde{\mathbf{h}}_{j,b,\mathcal{V}_{i}}^{\mathrm{H}},
\end{equation}
i.e., the sum in \eqref{eq:principal_submatrix} is restricted to same-BS UEs, and $Q_{k}$ and the safeguard rate count only same-cell streams. Terms with
$j\in\mathcal{K}_{b}\setminus\mathcal{N}_{b}(i)$ vanish on
the principal submatrix by \hyperref[lem:interference]{Lemma~\ref*{lem:interference}}.

\begin{proposition}[Decoupling under empty effective victim sets] \normalfont
\label{prop:decoupling}
Suppose $\mathcal{K}_{b}^{\mathrm{vic,eff}}=\emptyset$ for every
$b\in\mathcal{B}$. Then:
\begin{enumerate}
\item[(i)] $I_{k}^{\mathrm{inter}}=0$ for every $k\in\mathcal{K}$
and every beamformer set feasible for \eqref{prob:main}.
\item[(ii)] For every $b\in\mathcal{B}$ and every
$i\in\mathcal{K}_{b}$, $\mathcal{N}_{b}(i)\subseteq\mathcal{K}_{b}$,
so the multi-cell and per-cell update maps produce identical iterates from the same initialization when run with the same synchronized iteration and stopping schedule. Independently stopped cells need not return the identical final iterate.
\end{enumerate}
\end{proposition}

\noindent\textit{Proof:} see \hyperlink{proof:F}{Appendix~\ref*{app:proofs}}.

Under this common stopping schedule, coordination gains in the gated model require effective victims.

\subsection{Expected Loads and Effective Victims}
\label{sec:expected_loads}

The $\mathcal{L}_{b}^{\mathrm{VR}}$ identity below requires no independence assumption and holds for any association. For the $\mathcal{L}_{b}^{\mathrm{conf}}$ identity, assume \hypertarget{assumption:S1}{(S1)} an exogenous association fixed independently of the masks, \hypertarget{assumption:S2}{(S2)} independence across UEs, and \hypertarget{assumption:S3}{(S3)} arbitrary within-UE correlation. Expectations are over the masks.

\begin{proposition}[Expected VR-aware and conflict-aware loads]
\label{prop:expected_loads}
Under the assumptions above,
\begin{align}
 \mathbb{E}\!\left[\mathcal{L}_{b}^{\mathrm{VR}}\right]
 &=
 N_{\mathrm{s}}
 \sum_{k\in\mathcal{K}}\sum_{g\in\mathcal{G}}
 q_{k,b,g},
 \label{eq:exp_load_vr}
 \\
 \mathbb{E}\!\left[\mathcal{L}_{b}^{\mathrm{conf}}\right]
 &=
 N_{\mathrm{s}}
 \Bigg(
 \sum_{k\in\mathcal{K}_{b}}\sum_{g\in\mathcal{G}}
 q_{k,b,g}
 +
 \sum_{k\notin\mathcal{K}_{b}}\sum_{g\in\mathcal{G}}
 q_{k,b,g}\zeta_{b,g}
 \Bigg).
 \label{eq:exp_load_conf}
\end{align}
\end{proposition}

\noindent\textit{Proof:} see \hyperlink{proof:G}{Appendix~\ref*{app:proofs}}.

The conflict-load formula is a benchmark: gain-based association departs from \hyperlink{assumption:S1}{(S1)}, and shared blockers from \hyperlink{assumption:S2}{(S2)}. \hyperref[sec:replication]{Section~\ref*{sec:replication}} compares it with the measured loads.

\subsection{Robustness of VR-Net Under Soft Visibility}
\label{sec:robust}

The design uses the hard-gated channel $\tilde{\mathbf{h}}_{k,b}$, whereas the true channel is soft: a block outside $\mathcal{V}_{k,b}$ is attenuated, not removed. This subsection bounds the cost of the gating for any feasible beamformers of Problem~\eqref{prob:main} through the residual $\mathbf{r}_{k,b} = \mathbf{h}_{k,b} - \tilde{\mathbf{h}}_{k,b}$. The residual energy fraction is $\epsilon_{k,b} \triangleq \|\mathbf{r}_{k,b}\|^2/\|\mathbf{h}_{k,b}\|^2 \in [0,1]$ (set to zero for a zero channel).

\begin{lemma}[Exact desired signal]\normalfont
\label{lem:desired}
If $\operatorname{supp}_{\mathrm{sub}}(\mathbf{w}_k) \subseteq \mathcal{V}_{k,a_k}$, then
\begin{equation}
  \mathbf{h}_{k,a_k}^{H}\mathbf{w}_k
  = \tilde{\mathbf{h}}_{k,a_k}^{H}\mathbf{w}_k ,
\end{equation}
i.e., the desired signal is exact.
\end{lemma}

\begin{proof}
The residual $\mathbf r_{k,a_k}$ lies outside $\mathcal V_{k,a_k}$ and is orthogonal to the supported beamformer $\mathbf w_k$.

\end{proof}

\begin{lemma}[True interference, per term] \normalfont
\label{lem:interf}
Let $j \in \mathcal{K}_b$ with
$\operatorname{supp}_{\mathrm{sub}}(\mathbf{w}_j) \subseteq \mathcal{V}_{j,b}$. Then
\begin{equation}
  \mathbf{h}_{k,b}^{H}\mathbf{w}_j
  = \tilde{\mathbf{h}}_{k,b}^{H}\mathbf{w}_j
  + \mathbf{r}_{k,b}[\mathcal{V}_{j,b}]^{H}\mathbf{w}_j ,
\end{equation}
and
\begin{equation}
  \big|\mathbf{r}_{k,b}[\mathcal{V}_{j,b}]^{H}\mathbf{w}_j\big|
  \leq \big\|\mathbf{r}_{k,b}
  [\mathcal{V}_{j,b} \setminus \mathcal{V}_{k,b}]\big\|
  \,\|\mathbf{w}_j\| .
\end{equation}
If moreover $\mathcal{V}_{k,b} \cap \mathcal{V}_{j,b} = \emptyset$,
then $\tilde{\mathbf{h}}_{k,b}^{H}\mathbf{w}_j = 0$ and
\begin{equation}
  \big|\mathbf{h}_{k,b}^{H}\mathbf{w}_j\big|
  \leq \big\|\mathbf{h}_{k,b}[\mathcal{V}_{j,b}]\big\|
  \,\|\mathbf{w}_j\| .
\end{equation}
\end{lemma}

\noindent\textit{Proof:} see \hyperlink{proof:H}{Appendix~\ref*{app:proofs}}.

\begin{theorem}[Model-to-true SINR and rate gap] \normalfont
\label{thm:gap}
Let $\{\mathbf{w}_j\}$ be any feasible beamformers of
Problem~\eqref{prob:main}. For UE $k$, let
$S_k = |\mathbf{h}_{k,a_k}^{H}\mathbf{w}_k|^2$ be the
desired power, exact by \hyperref[lem:desired]{Lemma~\ref*{lem:desired}}.
Let $\hat{I}_k = \sum_{b}\sum_{j \in \mathcal{K}_b,\, j \neq k}
|\tilde{\mathbf{h}}_{k,b}^{H}\mathbf{w}_j|^2$ be the
interference on the design (gated) channel. Then
$\hat{\Gamma}_k = S_k / (\hat{I}_k + \sigma_k^2)$, and
let
\begin{equation}
  \Delta_k = \sum_{b}\sum_{j \in \mathcal{K}_b,\, j \neq k}
  \big\|\mathbf{r}_{k,b}
  [\mathcal{V}_{j,b} \setminus \mathcal{V}_{k,b}]\big\|^2
  \|\mathbf{w}_j\|^2 .
\end{equation}
\emph{(a) SINR bound:}
\begin{equation}
  \Gamma_k \;\geq\; \frac{S_k}{\big(\sqrt{\hat{I}_k}
  + \sqrt{\Delta_k}\big)^2 + \sigma_k^2} .
\end{equation}
\emph{(b) Rate gap:} with
$\hat{R}_k = \log_2(1 + \hat{\Gamma}_k)$ and
$R_k^{\mathrm{true}} = \log_2(1 + \Gamma_k)$,
\begin{equation}
  \hat{R}_k - R_k^{\mathrm{true}} \;\leq\;
  \log_2\!\left(1 + \frac{2\sqrt{\hat{I}_k \Delta_k} + \Delta_k}
  {\hat{I}_k + \sigma_k^2}\right) \;:=\; B_k ,
\end{equation}
and, with
$D_{k,-} = \big(\sqrt{\hat{I}_k} - \sqrt{\Delta_k}\big)_+^2
+ \sigma_k^2$,
\begin{equation}
  \hat{R}_k - R_k^{\mathrm{true}} \;\geq\;
  -\log_2\!\left(\frac{\hat{I}_k + \sigma_k^2}{D_{k,-}}\right) .
\end{equation}
For small $\Delta_k$ at fixed $\hat{I}_k$,
$B_k = \frac{2\sqrt{\hat{I}_k \Delta_k}}
{(\hat{I}_k + \sigma_k^2)\ln 2} + O(\Delta_k)$. The gap
is thus $O(\sqrt{\Delta_k})$. In the design-decoupled
case $\hat{I}_k = 0$, it reduces to
$B_k = \log_2(1 + \Delta_k/\sigma_k^2) = O(\Delta_k)$. With $x_k \triangleq \Delta_k/(\hat{I}_k+\sigma_k^2)$, $\hat{I}_k/(\hat{I}_k+\sigma_k^2)<1$ gives the universal envelope $B_k \leq 2\log_2(1+\sqrt{x_k})$.
\emph{(c) Approximate decoupling:} if $\mathcal{V}_{j,b} \cap \mathcal{V}_{k,b}
= \emptyset$ for every $b \neq a_k$ and every $j \in \mathcal{K}_b$
(the hypothesis of \hyperref[prop:decoupling]{Proposition~\ref*{prop:decoupling}} for UE $k$), then
$\hat{I}_k^{\mathrm{inter}} = 0$ and
\begin{equation}
\begin{split}
  I_k^{\mathrm{inter}}(\mathrm{true})
  &\leq \sum_{b \neq a_k}
  \big\|\mathbf{h}_{k,b}[\mathcal{U}_b]\big\|^2 P_b^{\max}\\
  &\leq \sum_{b \neq a_k} \epsilon_{k,b}\,
  \|\mathbf{h}_{k,b}\|^2 P_b^{\max} .
\end{split}
\end{equation}
\end{theorem}

\noindent\textit{Proof:} see \hyperlink{proof:I}{Appendix~\ref*{app:proofs}}.

\begin{remark}\label{rem:diagnostic} \normalfont
\hyperref[thm:gap]{Theorem~\ref*{thm:gap}} gives an a posteriori diagnostic, not a low-overhead runtime certificate. Computing its residual terms requires the out-of-VR true-channel blocks, which VR-Net does not upload. Since
$\|\mathbf{r}_{k,b}[\mathcal{V}_{j,b} \setminus \mathcal{V}_{k,b}]\|^2
\leq \|\mathbf{r}_{k,b}\|^2
= \epsilon_{k,b}\|\mathbf{h}_{k,b}\|^2$ and
$\sum_{j \in \mathcal{K}_b, j \neq k} \|\mathbf{w}_j\|^2
\leq P_b^{\max}$,
\begin{equation}
  \Delta_k \;\leq\; \sum_{b} \epsilon_{k,b}\,
  \|\mathbf{h}_{k,b}\|^2 P_b^{\max} .
\end{equation}
\end{remark}

\begin{theorem}[Support-constraint optimality gap] \normalfont
\label{thm:proj}
Fix the association. Let $\mathbf{W}^{\star} = \{\mathbf{w}_j^{\star}\}$ be any
feasible full-aperture beamformer set under the same per-BS power constraints. Here $\mathrm{WSR}_{\mathrm{VR,true}}^{\mathrm{opt}}$ is the largest true-channel WSR over the VR-supported beamformers. All WSR values use true channels, not design channels. Let $\mathbf{P}_j$ be the orthogonal
projection onto the coordinates of
$\mathcal{V}_{j,a_j}$, and let
\begin{equation}
  \psi_j = \frac{\|(\mathbf{I} - \mathbf{P}_j)
  \mathbf{w}_j^{\star}\|^2}{\|\mathbf{w}_j^{\star}\|^2}
\end{equation}
be the share of $\mathbf{w}_j^{\star}$ outside its VR ($\psi_j := 0$ if
$\mathbf{w}_j^{\star} = \mathbf{0}$). Let
$S_k^{\star}$, $I_k^{\star}$ and $\Gamma_k^{\star}$ be
the desired power, interference power and SINR of
$\mathbf{W}^{\star}$ on the true channel. Define
\begin{align}
  \Delta_k^{\epsilon} &= \epsilon_{k,a_k}\, \psi_k\,
  \|\mathbf{h}_{k,a_k}\|^2 \|\mathbf{w}_k^{\star}\|^2 , \\
  \Delta_k^{\psi} &= \sum_{\substack{b,\, j \in \mathcal{K}_b\\
  (b,j) \neq (a_k,k)}}
  \big\|\mathbf{h}_{k,b}[\mathcal{V}_{j,b}^{c}]\big\|^2 \, \psi_j \,
  \|\mathbf{w}_j^{\star}\|^2 .
\end{align}
Then $\mathbf{P}\mathbf{W}^{\star}$ is feasible for
Problem~\eqref{prob:main}. Its SINRs on the true channel
satisfy
\begin{equation}
  \Gamma_k(\mathbf{P}\mathbf{W}^{\star}) \;\geq\;
  \frac{\big(\sqrt{S_k^{\star}} -
  \sqrt{\Delta_k^{\epsilon}}\big)_+^2}
  {\big(\sqrt{I_k^{\star}} + \sqrt{\Delta_k^{\psi}}\big)^2
  + \sigma_k^2} .
\end{equation}
With $x_k = 2\sqrt{\Delta_k^{\epsilon}/S_k^{\star}}$
for $S_k^{\star} > 0$, the weighted sum rates satisfy
\begin{equation}
  \mathrm{WSR}_{\mathrm{VR,true}}^{\mathrm{opt}}
  \;\geq\; \mathrm{WSR}(\mathbf{P}\mathbf{W}^{\star})
  \;\geq\; \mathrm{WSR}(\mathbf{W}^{\star})
  - G(\epsilon, \psi) .
\end{equation}
Here
\begin{equation}
\begin{split}
  G(\epsilon, \psi) = \sum_{k} \omega_k \Big[&
  -\log_2\big(1 - x_k\big)\\
  &+ \log_2\!\Big(1 +
  \frac{2\sqrt{I_k^{\star}\Delta_k^{\psi}} + \Delta_k^{\psi}}
  {I_k^{\star} + \sigma_k^2}\Big) \Big] .
\end{split}
\end{equation}
The first term of $G$ is read as $0$ if
$S_k^{\star} = 0$. It is read as $+\infty$, hence
vacuous, if $x_k \geq 1$. For fixed channels and fixed
$\mathbf{W}^{\star}$, $G(\epsilon, \psi) \to 0$ as all
$\psi_j \to 0$, for any $\epsilon_{k,b} \in [0,1]$.

\end{theorem}

\noindent\textit{Proof:} see \hyperlink{proof:J}{Appendix~\ref*{app:proofs}}.

\begin{remark}\label{rem:projection} \normalfont
$G$ is not small in general: if the full-aperture solution relies on out-of-VR subarrays, e.g., for nulling, $G$ can be large or vacuous. Coarser forms of the two perturbation terms are
$\Delta_k^{\epsilon} \leq \epsilon_{k,a_k}\, \psi_k\,
\|\mathbf{h}_{k,a_k}\|^2 P_{a_k}^{\max}$ and
$\Delta_k^{\psi} \leq \big(\max_j \psi_j\big) \sum_b
\|\mathbf{h}_{k,b}\|^2 P_b^{\max}$. The latter uses
$\big\|\mathbf{h}_{k,b}[\mathcal{V}_{j,b}^{c}]\big\|
\leq \|\mathbf{h}_{k,b}\|$.
\end{remark}

\hyperref[thm:proj]{Theorem~\ref*{thm:proj}} bounds the true-channel VR-supported optimum relative to any feasible full-aperture reference at a fixed feasible association. It does not bound the output of \hyperref[alg:main]{Algorithm~\ref*{alg:main}}, which optimizes the gated channel; \hyperref[thm:gap]{Theorem~\ref*{thm:gap}} bounds that model mismatch; \hyperref[rem:diagnostic]{Remarks~\ref*{rem:diagnostic}} and~\hyperref[rem:projection]{\ref*{rem:projection}} clarify evaluation and nonvacuity.

\subsection{Conditions for Partial Visibility Regions}
\label{sec:exist}

A partial VR under \eqref{eq:vr_set} is determined by the relative-energy threshold and the INR gate. The geometric result below addresses the more specific single-edge LoS transition in which one block center has less than $1$~dB loss and another has more than $\eta$~dB loss; it is not a necessary condition for every operational partial VR. The NLoS power-floor result gives a separate necessary condition for a shadowed block to leave the relative-energy VR.

\begin{proposition}[Diffraction resolution] \normalfont
\label{prop:fresnel}
For this proposition, take $\eta>1$~dB, use the paraxial path expansion and assume that $L$ is monotone in $v$ across the transition; illuminated-side ripples are neglected. Consider a vertical knife edge whose tip lies on the BS--UE LoS ray, at distance $x$ from the array, with the UE at distance $d>x$. For an element at signed perpendicular offset $y$, let $L(y)$ be the single-edge loss of \hyperlink{cite.3gpp38901}{TR~38.901}, Sec.~7.6.4.2.
\emph{(a)} The loss depends on $y$ only through the
Fresnel parameter
\begin{equation}
  v(y) = y\,\sqrt{\frac{2(d - x)}{\lambda x d}} .
\end{equation}
The shadow transition is the set of $y$ where the loss
grows from $1$~dB to $\eta$~dB. Its width along the
aperture is
\begin{equation}
  w_F(x) = c\,\sqrt{\frac{\lambda x (d - x)}{d}}\,
  \frac{d}{d - x},
  \qquad
  c = \frac{v_{\eta} - v_1}{\sqrt{2}} .
\end{equation}
Here $v_1$ and $v_{\eta}$ are the Fresnel parameters at
which the loss equals $1$~dB and $\eta$~dB. Also,
$v_1 < 0$, and $v_{\eta} \gtrless 0$ according as
$\eta \gtrless 6.02$~dB, the grazing loss. Inverting $L = -20\log_{10}(0.5 - F)$ gives the parameter at loss $L$~dB in closed form,
$v_L = \tan\big(\pi\,(0.5 - 10^{-L/20})\big)\big/
\big(\pi^{3/2}/4\big)$; e.g., $c(10~\mathrm{dB})=1.759$.

\emph{(b)} Let block losses be evaluated at block
centers. Let $L_{\mathrm{block}}$ be the projected block
length along the perpendicular direction. Let
$D_{\mathrm{ap}}$ be the aperture span projected
perpendicular to the ray. A necessary condition for one
block to be shadowed by more than $\eta$~dB, while
another is shadowed by less than $1$~dB, is
\begin{equation}
  D_{\mathrm{ap}} > w_F(x) + L_{\mathrm{block}} .
\end{equation}
\end{proposition}

\noindent\textit{Proof:} see \hyperlink{proof:K}{Appendix~\ref*{app:proofs}}.

If every element of one block has loss above $\eta$~dB and every element of another has loss below $1$~dB, the same argument gives $D_{\mathrm{ap}} > w_F(x) + 2L_{\mathrm{block}}$.

\begin{proposition}[NLoS power floor] \normalfont
\label{prop:floor}
Consider block powers $P_g=P_g^{\mathrm{LoS}}+P_g^{\mathrm{NLoS}}$ with strictly positive NLoS powers and $\max_g P_g^{\mathrm{NLoS}}/\min_g P_g^{\mathrm{NLoS}}\leq\rho$, $\rho\geq1$. Let $g^\star$ be a strongest block and $g_0$ any shadowed block, including one with nonzero residual LoS power. Define $\mathcal K_{g^\star}\triangleq P_{g^\star}^{\mathrm{LoS}}/P_{g^\star}^{\mathrm{NLoS}}$. Then
\begin{equation}
  \frac{P_{g_0}}{P_{g^\star}}\geq\frac{1}{\rho(1+\mathcal K_{g^\star})}.
\end{equation}
Thus, a necessary condition for $g_0$ to fall below the relative $\eta$-dB threshold is $10\log_{10}[\rho(1+\mathcal K_{g^\star})]>\eta$. The INR gate in \eqref{eq:vr_set} can exclude a block independently of this condition.
\end{proposition}
\begin{proof}
Since $P_{g_0}\geq P_{g_0}^{\mathrm{NLoS}}\geq P_{g^\star}^{\mathrm{NLoS}}/\rho$ and $P_{g^\star}=(1+\mathcal K_{g^\star})P_{g^\star}^{\mathrm{NLoS}}$, the bound follows. Exclusion by the relative threshold requires $P_{g_0}/P_{g^\star}<10^{-\eta/10}$, giving the necessary condition.
\end{proof}
For a single LoS-blocking edge, the contrast in \hyperref[prop:fresnel]{Proposition~\ref*{prop:fresnel}} requires sufficient aperture span, while the NLoS floor in \hyperref[prop:floor]{Proposition~\ref*{prop:floor}} limits the achievable energy contrast. Operational VRs use both gates in \eqref{eq:vr_set}.

\section{Numerical Results}
\label{sec:results}

\subsection{Simulation Setup}
\label{sec:setup}

At $15$~GHz, the simulator uses a customized Release~19-based UMi Street Canyon channel. It adopts the \hyperlink{cite.3gpp38901}{TR~38.901} path-loss and LoS-probability parameters, UMi cluster-delay/power statistics, azimuth-ray offsets, per-element antenna pattern, and physical-blocker SNS. Its spherical-wave element responses follow \cite{xu2026}; departures from the full Release~19 procedure are listed below and in \hyperref[sec:channel]{Section~\ref*{sec:channel}} (\hyperref[tab:setup]{Table~\ref*{tab:setup}}). The reported simulator evaluates the knife-edge excess path geometrically; \hyperref[prop:fresnel]{Proposition~\ref*{prop:fresnel}} uses its paraxial width approximation. The sweeps vary one quantity at a time: transmit power ($20$, $27$, $40$~dBm), UEs per cell ($4$, $12$) and module gap ($0$, $0.5$, $2.0$~m).

\begin{table}[t]
\caption{Main simulation setting.}
\label{tab:setup}
\centering
\footnotesize
\begin{tabular}{@{}p{0.27\columnwidth}@{\hspace{4pt}}p{0.69\columnwidth}@{}}
\hline
\raggedright Parameter & \raggedright Value \tabularnewline
\hline
\raggedright Carrier, noise bandwidth & \raggedright $15$~GHz ($\lambda = 2$~cm), $100$~MHz \tabularnewline
\raggedright Noise & \raggedright $-174$~dBm/Hz, NF $7$~dB ($-87$~dBm) \tabularnewline
\raggedright Deployment & \raggedright $B = 3$ BSs, ISD $200$~m, heights $10$/$1.5$~m \tabularnewline
\raggedright UEs & \raggedright $8$ per cell ($K=24$), outdoor; angle uniform in $\pm 60^{\circ}$, distance $\sqrt{\cdot}$-uniform in $[10, 115.5]$~m \tabularnewline
\raggedright Array per BS & \raggedright $N = 256$: $G = 16$ modules of $N_{\mathrm{s}} = 16$, $\lambda/2$ spacing, gap $1.0$~m, span $17.55$~m \tabularnewline
\raggedright Element pattern & \raggedright \hyperlink{cite.3gpp38901}{TR~38.901} Table~7.3-1, per element \tabularnewline
\raggedright Transmit power & \raggedright $33$~dBm per BS \tabularnewline
\raggedright Large scale & \raggedright UMi-Street Canyon path loss, shadowing and LoS probability (\hyperlink{cite.3gpp38901}{TR~38.901} Tables~7.4.1-1, 7.4.2-1) \tabularnewline
\raggedright Clusters & \raggedright $12$ (LoS) / $19$ (NLoS), $20$ rays, $-25$~dB pruning; near-field sources as in \hyperref[sec:system_model]{Section~\ref*{sec:system_model}} \tabularnewline
\raggedright SNS & \raggedright physical blockers (Sec.~7.6.14.1.2), scenario C, level high; knife-edge loss per element and ray \tabularnewline
\raggedright Blockers & \raggedright Poisson, $4$ per $1000$~m$^2$ in $5$--$60$~m: billboard $20\%$ ($2.4 \times 3.6$~m), lamp $50\%$ ($0.4 \times 0.8$~m), vehicle $30\%$ ($4.8 \times 1.4$~m); per-ray screen rotation; one building edge per BS at $10$--$50$~m \tabularnewline
\raggedright VR rule & \raggedright threshold $\eta$, INR gate $\gamma_{\min} = -10$~dB \tabularnewline
\raggedright Beamforming & \raggedright WMMSE, $\leq 60$ iterations, tolerance $10^{-4}$, safeguard \tabularnewline
\raggedright CSI & \raggedright perfect on the blocks each scheme uses \tabularnewline
\hline
\end{tabular}
\end{table}

The $100$~MHz bandwidth sets noise power; rates use a frequency-flat channel. The $17.55$~m aperture permits up to $58.5$~ns geometric delay spread before multipath, so wideband performance requires separate evaluation.

The modelling choices relative to \hyperlink{cite.3gpp38901}{TR~38.901} are: one polarization; independent large-scale parameters; cluster ZOD equal to the LoS ZOD plus $\mathcal N(0,3^\circ)$, without per-ray ZOD spread or LoS delay scaling by $C_\tau$; an isotropic scalar UE, so arrival angles are not needed; source distances and delays as in \hyperref[sec:channel]{Section~\ref*{sec:channel}}; each blocker screen oriented perpendicular to the ray it intercepts; and, for NLoS rays, knife-edge loss evaluated on the finite scatterer-to-element leg.

\emph{Schemes.} VR-Net (\hyperref[sec:algorithm]{Section~\ref*{sec:algorithm}}) runs at $\eta \in \{10, 15, 20, 30\}$~dB; all schemes are scored by the true rates \eqref{eq:sinr_true}. The comparison schemes are:
\begin{itemize}
\item \emph{Per-cell VR-aware:} the per-cell variant of
\hyperref[sec:decoupling]{Section~\ref*{sec:decoupling}} at the same $\eta$.
\item \emph{Full-aperture WMMSE:} the same iteration without gating, full CSI at every BS.
\item \emph{Adapted P\'erez-Ad\'an et al.\ \cite{perezadan2025}, full CSI:} their user selection and multi-cell zero-forcing on the full channel, with the original stopping rule removed so that every candidate user is served.
\item \emph{Adapted P\'erez-Ad\'an et al., VR-gated CSI:} the same adapted all-user procedure on $\tilde{\mathbf{h}}_{k,b}$, at the same threshold. It uses all VR-gated blocks ($\mathcal{L}_{b}^{\mathrm{VR}}$), whereas VR-Net uploads only \eqref{eq:conflict_blocks}, so it receives at least as much CSI as VR-Net.
\item \emph{Dual-threshold VR-Net (ablation):} supports from a $10$~dB mask, interference terms from a wider $\eta_{\mathrm{d}}$ mask.
\end{itemize}
A UE with an empty VR at every BS is excluded from all schemes. Emptiness depends only on the INR gate, so the population is common to all thresholds; where recorded, exclusions are rare: $6$ of $720$ UE slots at $20$~dBm, none at $27$ or $40$~dBm, and none in the $1200$ UE slots of Set~1.

\emph{Drops and statistics.} A drop is one realization of UE positions, blockers and channels. Set~1 ($50$ drops, seeds $7$/$107$) served development and diagnostics. Set~2 ($50$ drops, seeds $8$/$108$), drawn after its criteria were fixed in writing, carries the main comparison. Set~3 ($30$ drops, seeds $9$/$109$) carries the replication and the sweeps. Intervals are $95\%$ percentile bootstrap intervals over drops ($5000$ resamples), paired per drop for differences. Rate retention is the ratio of mean sum rates; CSI load is the mean per-drop load ratio. UE percentiles pool eligible UE--drop rates.

Bootstrap intervals are pointwise. Comparisons with \cite{perezadan2025} concern its all-user adaptation, whose scheduling objective differs from the original.

\subsection{Main Comparison on the Evaluation Drops}
\label{sec:main_results}

\begin{figure}[t]
\centering
\includegraphics[width=\columnwidth]{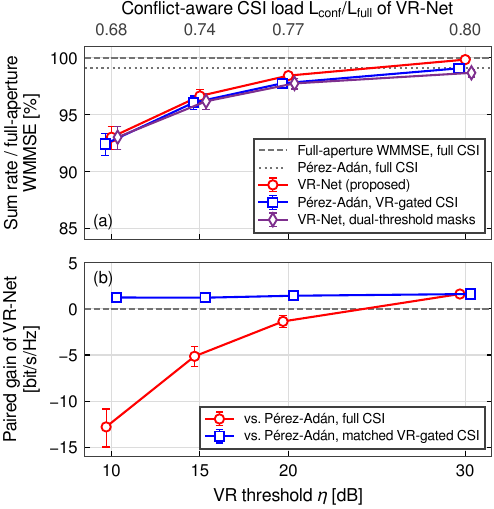}
\caption{Set~2, $50$ drops. (a) Sum rate relative to full-aperture WMMSE against $\eta$; top axis: conflict-aware CSI load of VR-Net. (b) Paired sum-rate gain of VR-Net over the all-user adaptation of \cite{perezadan2025} with full and with VR-gated CSI. Bars: $95\%$ paired bootstrap intervals.}
\label{fig:headline}
\end{figure}

\begin{figure}[t]
\centering
\includegraphics[width=\columnwidth]{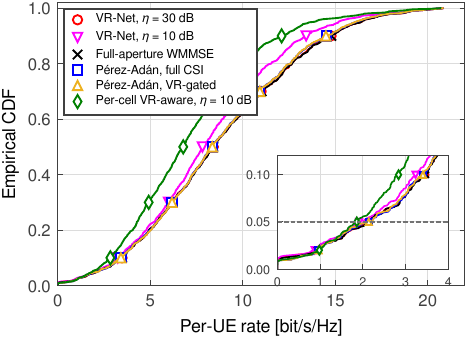}
\caption{Empirical CDF of the per-UE true rate, Set~2 ($1200$ UE slots). Inset: lower tail; dashed line: $5$th percentile.}
\label{fig:ratecdf}
\end{figure}

\hyperref[tab:main_results]{Table~\ref*{tab:main_results}} and \hyperref[fig:headline]{Fig.~\ref*{fig:headline}} report Set~2. The VR-Net sum rate rises with $\eta$, from $93.0\%$ of full-aperture WMMSE at $\eta=10$~dB to $99.8\%$ at $\eta=30$~dB, while the conflict-aware load rises from $0.676$ to $0.796$ of the full load. In the VR-gated comparison, VR-Net outperforms the VR-gated all-user adaptation of \cite{perezadan2025} at every $\eta$, by $+1.22$ to $+1.62$~bit/s/Hz. At $\eta=30$~dB it also outperforms the full-CSI all-user adaptation of \cite{perezadan2025}, by $+1.62$ $[+1.24,+2.00]$~bit/s/Hz, with $20\%$ less CSI. At $\eta=10$~dB, VR-Net exceeds the per-cell VR-aware rate of $172.07$~bit/s/Hz by $24.56$ $[20.86,28.42]$~bit/s/Hz, a $14.3\%$ gain relative to per-cell design. \hyperref[fig:ratecdf]{Fig.~\ref*{fig:ratecdf}} shows the per-UE rate distributions. At $\eta=30$~dB, the largest vertical distance between the CDF of VR-Net and those of full-aperture WMMSE and the two adaptations of \cite{perezadan2025} is $0.006$ and $0.018$, against $0.18$ for per-cell design; the medians are $8.40$, $8.41$ and $8.35$~bit/s/Hz. Per-cell design shifts the distribution left (median $6.78$~bit/s/Hz). The pooled $5$th-percentile rate of VR-Net is $1.92$ and $2.11$~bit/s/Hz at $\eta=10$ and $30$~dB, against $2.16$ for full-aperture WMMSE and $1.88$ for per-cell design. At $\eta=30$~dB, VR-Net gives a positive rate to $23.8$ of the $24$ UEs per drop on average; the remaining $0.9\%$ of UE slots appear at zero rate in \hyperref[fig:ratecdf]{Fig.~\ref*{fig:ratecdf}}. \hyperref[tab:main_results]{Table~\ref*{tab:main_results}} lists all sixteen pre-registered criteria: eleven pass and five fail.

\begin{table*}[t]
\footnotesize
\caption{Main results and pre-registered criteria. Mean sum rates in bit/s/Hz, with the ratio to full-aperture WMMSE in parentheses. Paired differences with $95\%$ bootstrap intervals. Load: mean $\mathcal{L}_{\mathrm{conf}}/\mathcal{L}_{\mathrm{full}}$ unless stated. C1--C16 identify the criteria; superiority requires a positive paired interval, C2 additionally requires $10\%$ gain, and p5 denotes the pooled fifth-percentile ratio to full-aperture WMMSE.}
\label{tab:main_results}
\centering
\setlength{\tabcolsep}{3.5pt}\begin{tabular}{@{}p{0.18\textwidth}lp{0.30\textwidth}lp{0.25\textwidth}@{}}
\hline
\raggedright Scheme & Mean rate & \raggedright Paired difference & Load & \raggedright Criterion outcome \tabularnewline
\hline
\raggedright Full-aperture WMMSE (Set~2) & $211.42$ & \raggedright reference & $1.000$ & \raggedright n/a \tabularnewline
\raggedright P\'erez-Ad\'an et al. \cite{perezadan2025}, full CSI (Set~2) & $209.44$ & \raggedright reference & $1.000$ & \raggedright n/a \tabularnewline
\raggedright VR-Net, $\eta = 10$~dB (Set~2) & $196.64$ ($0.930$) & \raggedright $+1.24$ $[+0.68, +1.81]$ vs VR-gated & $0.676$ & \raggedright C16 component (see $\eta=30$ row) \tabularnewline
\raggedright VR-Net, $\eta = 15$~dB (Set~2) & $204.30$ ($0.966$) & \raggedright $+1.22$ $[+0.80, +1.63]$ vs VR-gated & $0.737$ & \raggedright C16 component (see $\eta=30$ row) \tabularnewline
\raggedright VR-Net, $\eta = 20$~dB (Set~2) & $208.09$ ($0.984$) & \raggedright $+1.44$ $[+1.05, +1.80]$ vs VR-gated & $0.767$ & \raggedright C15, $\geq98\%$ of full-ap.: PASS \tabularnewline
\raggedright VR-Net, $\eta = 30$~dB (Set~2) & $211.06$ ($0.998$) & \raggedright $+1.62$ $[+1.28, +1.95]$ vs VR-gated; $+1.62$ $[+1.24, +2.00]$ vs full CSI & $0.796$ & \raggedright C13, full CSI: PASS; C14, load $\leq0.85$: PASS; C16, VR-gated at all four thresholds: PASS \tabularnewline
\raggedright Dual-threshold, $\eta_{\mathrm{d}} = 20$~dB (Set~2) & $206.64$ ($0.977$) & \raggedright $-0.01$ $[-0.69, +0.59]$ vs dual-gated; $+28.30$ $[+24.14, +32.55]$ vs per-cell dual; $+10.00$ $[+8.43,+11.67]$ vs single $\eta=10$ & $0.760$ & \raggedright C7, $\geq97\%$ of full-ap.: PASS; C8, load $\leq0.85$: PASS; C9, single $\eta=10$: PASS; C10, dual-gated: FAIL; C11, p5 $0.891<0.90$: FAIL; C12, per-cell dual: PASS \tabularnewline
\raggedright VR-Net, $\eta = 10$~dB (Set~1) & $197.62$ ($0.934$) & \raggedright $-12.55$ $[-14.80, -10.39]$ vs full CSI; $+0.75$ $[+0.34, +1.19]$ vs VR-gated; $+24.93$ $[+21.45, +28.50]$ vs per-cell & $0.751$ (of $\mathcal{L}_{\mathrm{UE}}$) & \raggedright C1, $\geq95\%$ of full-ap.: FAIL; C2, per-cell $+10\%$: PASS; C3, load $\leq0.75$: FAIL; C4, p5 $\geq0.90$: PASS; C5, full CSI: FAIL; C6, VR-gated: PASS \tabularnewline
\hline
\end{tabular}
\end{table*}

\begin{figure*}[t]
\centering
\begin{minipage}[t]{0.48\textwidth}
\centering
\includegraphics[width=\linewidth]{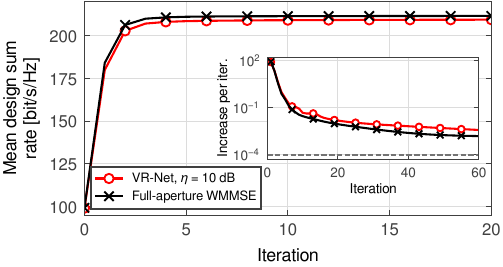}
\par\smallskip (a)
\end{minipage}\hfill
\begin{minipage}[t]{0.48\textwidth}
\centering
\includegraphics[width=\linewidth]{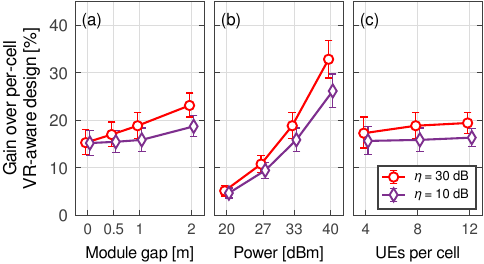}
\par\smallskip (b)
\end{minipage}
\caption{Set~1 convergence and Set~3 coordination. (a) Mean design sum rate per iteration at $\eta=10$~dB; inset: mean increase and the $10^{-4}$ tolerance. (b) Relative sum-rate gain of VR-Net over per-cell VR-aware design against module gap, transmit power and UEs per cell, with $95\%$ bootstrap intervals.}
\label{fig:conv_coord}
\end{figure*}

\hyperref[fig:conv_coord]{Fig.~\ref*{fig:conv_coord}(a)} shows the safeguarded iteration on Set~1. The mean design sum rate is monotone and reaches $99.9\%$ of its final value within $27$ iterations, against $14$ for full-aperture WMMSE. The $10^{-4}$ tolerance is rarely met before the $60$-iteration cap, so the cap, not the tolerance, sets the complexity.

\begin{figure*}[t]
\centering
\begin{minipage}[t]{0.48\textwidth}
\centering
\includegraphics[width=\linewidth]{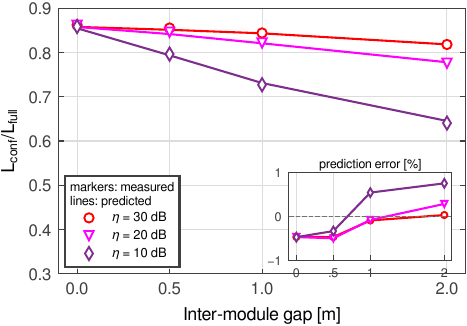}
\par\smallskip (a)
\end{minipage}\hfill
\begin{minipage}[t]{0.48\textwidth}
\centering
\includegraphics[width=\linewidth]{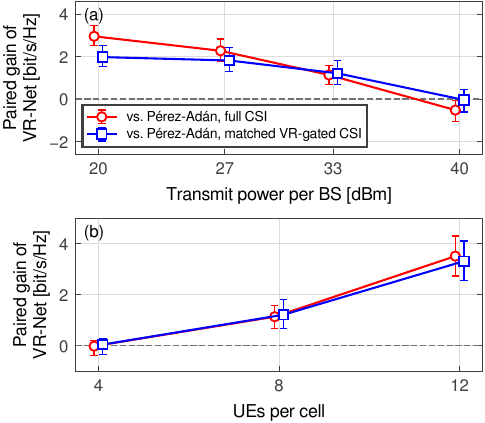}
\par\smallskip (b)
\end{minipage}
\caption{CSI loads and parameter sweeps on Set~3. (a) Conflict-aware load against the module gap: measurements (markers) and \hyperref[prop:expected_loads]{Proposition~\ref*{prop:expected_loads}} with empirical visibility frequencies (lines); inset: relative prediction error. (b) Paired sum-rate gain of VR-Net ($\eta=30$~dB) over the all-user adaptation of \cite{perezadan2025} with full and with VR-gated CSI against the transmit power per BS (top) and the UEs per cell (bottom); bars: $95\%$ paired bootstrap intervals.}
\label{fig:loads}
\label{fig:sweep}
\end{figure*}

\begin{figure*}[t]
\centering
\includegraphics[width=\textwidth]{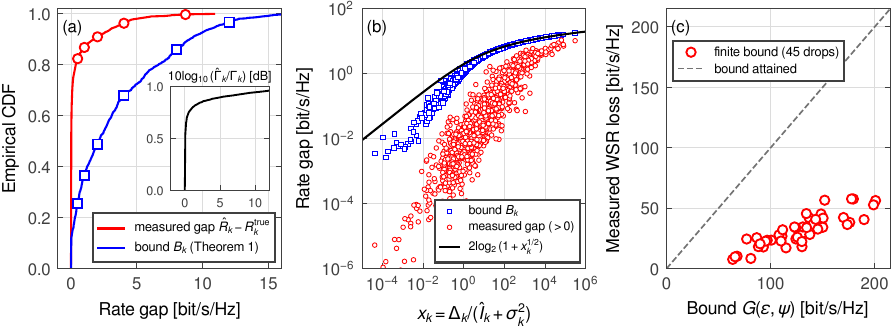}
\caption{Bounds on Set~1 ($\eta=10$~dB, $1200$ UE slots). (a) Empirical CDFs of the design-to-true rate gap and the bound $B_k$ of \hyperref[thm:gap]{Theorem~\ref*{thm:gap}}; inset: CDF of the design-to-true SINR ratio. (b) Bound and positive gaps against $x_k=\Delta_k/(\hat{I}_k+\sigma_k^2)$, with the envelope $2\log_2(1+\sqrt{x_k})$. (c) Measured weighted-sum-rate loss of the projected full-aperture solution against the bound $G$ of \hyperref[thm:proj]{Theorem~\ref*{thm:proj}} for the $45$ drops with a finite bound.}
\label{fig:theorems}
\end{figure*}

\begin{figure*}[t]
\centering
\begin{minipage}[t]{0.48\textwidth}
\centering
\includegraphics[width=\linewidth]{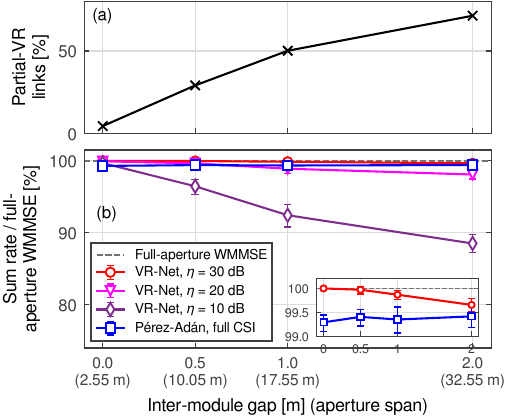}
\par\smallskip (a)
\end{minipage}\hfill
\begin{minipage}[t]{0.48\textwidth}
\centering
\includegraphics[width=\linewidth]{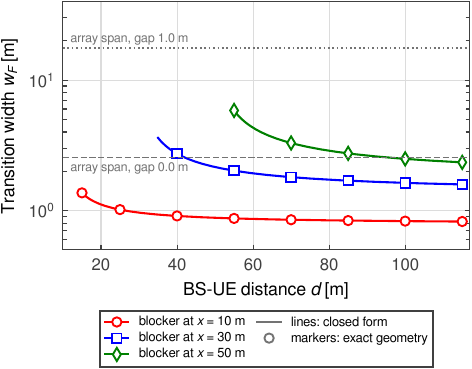}
\par\smallskip (b)
\end{minipage}
\caption{Module-gap and Fresnel analysis on Set~3. (a) Partial-VR share among nonempty links at $\eta=10$~dB (top) and sum rate relative to full-aperture WMMSE (bottom; inset: zoom near $100\%$). (b) Fresnel transition width $w_F$ at $15$~GHz against BS--UE distance for single edges at $x=10$, $30$ and $50$~m; horizontal lines show the contiguous and modular array spans. Bars in (a): $95\%$ bootstrap intervals.}
\label{fig:gap_fresnel}
\end{figure*}

\subsection{Replication and Gap Decomposition}
\label{sec:replication}
\label{sec:gap}

Set~3 reports a similar comparison. At $\eta=30$~dB, VR-Net reaches $219.21$~bit/s/Hz, $99.87\%$ of full-aperture WMMSE, at load $0.844$. It outperforms the full-CSI all-user adaptation of \cite{perezadan2025} by $+1.14$ $[+0.67,+1.58]$ and the VR-gated scheme by $+1.21$ $[+0.68,+1.81]$~bit/s/Hz. The per-cell coordination gain is $+27.78$ $[+23.30,+32.35]$ and $+34.74$ $[+29.84,+39.69]$~bit/s/Hz at $\eta=10$ and $30$~dB. At $\eta=10$~dB, VR-Net reaches $92.4\%$ of full-aperture WMMSE on Set~3 and $93.4\%$ on Set~1. \hyperref[fig:loads]{Fig.~\ref*{fig:loads}(a)} checks \hyperref[prop:expected_loads]{Proposition~\ref*{prop:expected_loads}} with the empirical visibility frequencies of each UE slot. The prediction of $\mathcal{L}_{\mathrm{VR}}$ is exact. The prediction of $\mathcal{L}_{\mathrm{conf}}$ lies within $0.75\%$ of the measured mean in all twelve gap-threshold settings of Set~3 and within $0.73\%$ on Set~1, although gain-based association and shared blockers depart from \hyperlink{assumption:S1}{(S1)}--\hyperlink{assumption:S2}{(S2)}. The error changes sign with the gap: it is negative for contiguous arrays and positive for wide gaps at $\eta=10$~dB.

\hyperref[tab:gap]{Table~\ref*{tab:gap}} splits the Set~1 gap at $\eta=10$~dB with a numerical reference that redesigns the beamformers on the true channel over the same supports. The empirical full-aperture-to-supported-reference difference is about $15\%$ of the gap, and the supported-reference-to-VR-Net difference is about $85\%$. These differences also include local-solver and initialization effects and do not identify the global support and mismatch losses. Raising $\eta$ changes both support and design channels: the ratio to full-aperture WMMSE rises to $0.971$, $0.989$ and $0.999$ at $15$, $20$ and $30$~dB. The mean ratio of residual-only interference power to true interference power is $45.3\%$, averaged over UE--drop pairs with nonzero true interference. This ratio excludes desired power and noise and need not be bounded by one because residual and gated interference can cancel.

\begin{table}[t]
\footnotesize
\caption{Gap to full-aperture WMMSE, Set~1, $\eta = 10$~dB. Mean sum rates in bit/s/Hz; losses paired per drop, $95\%$ intervals.}
\label{tab:gap}
\centering
\begin{tabular}{@{}p{0.63\columnwidth}r@{}}
\hline
Quantity & Value \\
\hline
Full-aperture WMMSE & $211.70$ \\
True-channel numerical redesign on VR supports & $209.56$ \\
VR-Net & $197.62$ \\
Full-to-supported reference difference & $2.13$ $[1.76, 2.54]$ \\
Supported-reference-to-VR-Net difference & $11.94$ $[9.95, 14.06]$ \\
Total gap & $14.07$ $[11.90, 16.35]$ \\
\hline
\end{tabular}
\end{table}

\hyperref[fig:theorems]{Fig.~\ref*{fig:theorems}} evaluates both theorems on Set~1. Both rate bounds of \hyperref[thm:gap]{Theorem~\ref*{thm:gap}} hold for all $1200$ eligible UE slots. The measured gap is below $0.014$~bit/s/Hz for half of the UEs and below $1.71$~bit/s/Hz for $90\%$; it is zero for $131$ UEs and negative for $78$, where residual and gated interference partly cancel. The design SINR overstates the true SINR by a median of $0.05$~dB and by $5.23$~dB at the $90$th percentile (inset). Every bound lies below the envelope $2\log_2(1+\sqrt{x_k})$, and for $x_k<1$ the positive gaps lie a median of $2.7$ decades below it; the median gap-to-bound ratio is $0.014$. \hyperref[thm:proj]{Theorem~\ref*{thm:proj}} is finite in $45$ of $50$ drops and holds in all of them: the measured loss of projecting full-aperture WMMSE onto the VRs is $7.7$ to $57.7$~bit/s/Hz, against bounds of $63.6$ to $200.9$~bit/s/Hz (median ratio $4.1$). In the $5$ vacuous drops, the measured loss is $21.1$ to $35.6$~bit/s/Hz.

\emph{Ablations.} The dual-threshold ablation lies below the single-threshold frontier at comparable loads (\hyperref[tab:main_results]{Table~\ref*{tab:main_results}}); its VR-gated difference is inconclusive. A surrogate local search over the association changes the Set~1 sum rate by $+0.26$ $[-0.11,+0.60]$~bit/s/Hz, so the gain-based initialization is kept.

\subsection{Parameter Sweeps}
\label{sec:sweeps}

Each sweep varies one parameter of the main setting on the $30$ drops of Set~3.

\emph{Module gap.} \hyperref[fig:gap_fresnel]{Fig.~\ref*{fig:gap_fresnel}(a)} shows the trend suggested by \hyperref[sec:exist]{Section~\ref*{sec:exist}}. At $\eta=10$~dB, the share of nonempty links with a partial VR grows from $4.5\%$ for the contiguous array ($2.55$~m span) to $29.2\%$, $50.1\%$ and $71.4\%$ for gaps of $0.5$, $1.0$ and $2.0$~m. The CSI saving grows with it (\hyperref[fig:loads]{Fig.~\ref*{fig:loads}(a)}): $\mathcal{L}_{\mathrm{conf}}/\mathcal{L}_{\mathrm{full}}$ falls from $0.86$ to $0.82$ at $\eta=30$~dB and from $0.86$ to $0.64$ at $\eta=10$~dB. At $\eta=30$~dB, VR-Net keeps at least $99.6\%$ of full-aperture WMMSE at every gap and outperforms the full-CSI all-user adaptation of \cite{perezadan2025} at every gap; the margin narrows from $+1.48$ $[+1.20,+1.84]$ to $+0.55$ $[+0.04,+1.00]$~bit/s/Hz at $2.0$~m, where the shortfall of $\eta=10$~dB is largest ($88.5\%$). \hyperref[fig:gap_fresnel]{Fig.~\ref*{fig:gap_fresnel}(b)} illustrates the geometric mechanism of \hyperref[prop:fresnel]{Proposition~\ref*{prop:fresnel}}. For edges $10$--$50$~m from the array, the transition width lies between $0.8$ and $5.8$~m: comparable to the contiguous span, but well below the modular span. On the plotted grid, with the UE at least $5$~m beyond the edge, the closed form agrees with the exact knife-edge geometry within $0.27\%$.

\emph{Power and load.} \hyperref[fig:sweep]{Fig.~\ref*{fig:sweep}(b)} compares VR-Net with the all-user adaptation of \cite{perezadan2025}. At $\eta=30$~dB, VR-Net outperforms the full-CSI scheme by $+2.96$, $+2.28$ and $+1.14$~bit/s/Hz at $20$, $27$ and $33$~dBm, and the VR-gated scheme by $+1.98$, $+1.82$ and $+1.21$~bit/s/Hz, with $39\%$, $24\%$ and $16\%$ less CSI than full exchange. At $40$~dBm the VR-gated comparison is inconclusive ($-0.04$ $[-0.61,+0.44]$), while VR-Net trails the full-CSI scheme by $0.52$ $[0.07,1.06]$~bit/s/Hz ($0.19\%$) while exchanging $10\%$ less CSI. There, the network is interference-limited and full-aperture zero-forcing approaches the WMMSE numerical reference ($99.92\%$). A pre-registered follow-up restores the advantage: on $30$ fresh drops (seeds $10$/$110$) at $40$~dBm, VR-Net at $\eta=40$~dB reaches $99.94\%$ of full-aperture WMMSE and outperforms the full-CSI scheme by $+0.47$ $[+0.02,+0.89]$~bit/s/Hz with $12\%$ less CSI; in the VR-gated comparison the difference is $+0.37$ $[-0.09,+0.80]$~bit/s/Hz. The advantage grows with user density: $-0.02$ $[-0.39,+0.21]$, $+1.14$ and $+3.52$ $[+2.74,+4.31]$~bit/s/Hz at $4$, $8$ and $12$ UEs per cell, with the same ordering in the VR-gated comparison. With $4$ UEs per cell, both schemes reach at least $99.9\%$ of full-aperture WMMSE. Across the power and load points, VR-Net retains at least $99.7\%$ of full-aperture WMMSE at $\eta=30$~dB.

\emph{Coordination gain.} \hyperref[fig:conv_coord]{Fig.~\ref*{fig:conv_coord}(b)} isolates the value of inter-cell coordination: the gain of VR-Net over per-cell VR-aware design at the same $\eta$. At $\eta=30$~dB it grows with the gap, from $15.3\%$ to $23.1\%$, and with the power, from $5.1\%$ at $20$~dBm to $32.8\%$ at $40$~dBm, and lies between $17.3\%$ and $19.4\%$ across user densities. 
\FloatBarrier
\section{Conclusion}\label{sec:conclusion}
VR-Net coordinates multi-cell XL-MIMO beamforming through BS-indexed VR overlaps and uploads only conflict-relevant channel blocks. The analysis establishes decoupling conditions, CSI loads, and separate bounds for the support and gated-channel losses; both rate bounds held for every evaluated UE. At a $30$-dB threshold, VR-Net retained at least $99.6\%$ of full-aperture WMMSE with fewer CSI entries in every setting. It outperformed the all-user adaptation of P\'erez-Ad\'an et al.\ at moderate power and dense load and stayed within $0.2\%$ of it at high power or light load, where a $40$-dB threshold restored a significant advantage. The CSI saving grows with the module gap, as the partial-VR conditions predict. Future work includes gated-block CSI acquisition, wideband channels, multi-antenna UEs, and joint transmission.

\appendices
\section{Proofs}\label{app:proofs}
\hypertarget{proof:A}{}\begin{proof}[{\proofref{lem:rank}{A. Rank of the VR-gated channel covariance}}]\leavevmode\\*
Conditional on the large-scale state in \hyperlink{assumption:A5}{(A5)}, $\mathbf R_{k,b}=\frac{\varrho_{k,b}}{1+\mathcal K_{k,b}}\sum_{c,\ell}\sigma_{k,b,c,\ell}^2
(\mathbf D_{k,b}\mathbf a_{k,b}^{(c,\ell)})(\mathbf D_{k,b}\mathbf a_{k,b}^{(c,\ell)})^{\mathrm H}$.
Each term has rank one and is supported on the $|\mathcal V_{k,b}|N_{\mathrm s}$
visible coordinates. The sum therefore has zero rows and columns outside the VR,
and rank at most both $\sum_cL_{k,b,c}$ and $|\mathcal V_{k,b}|N_{\mathrm s}$.
\end{proof}

\hypertarget{proof:B}{}\begin{proof}[{\proofref{lem:interference}{B. Interference bound scaled by subarray intersection}}]\leavevmode\\*
Only blocks $g\in\mathcal I_{k,j,b}$ contribute to
$\tilde{\mathbf h}_{k,b}^{\mathrm H}\mathbf w_{j,b}$.
Cauchy--Schwarz over the stacked blocks gives \eqref{eq:interference_bound_1}.
Applying $\|\mathbf u+\mathbf v\|^2\leq2\|\mathbf u\|^2+2\|\mathbf v\|^2$
and the maximum-entry definition of $A_{k,b,g}^{\mathrm{LoS}}$ gives
\eqref{eq:interference_bound_2}; bounding each summand by its maximum gives
\eqref{eq:interference_bound_3}. An empty intersection gives zero.
\end{proof}

\hypertarget{proof:C}{}\begin{proof}[{\proofref{prop:wmmse_convergence}{C. Monotonicity of the safeguarded iteration}}]\leavevmode\\*
The safeguard accepts only non-decreasing design WSR values. Moreover,
$\gamma_k\leq\|\tilde{\mathbf h}_{k,a_k}\|^2P_{a_k}^{\max}/\sigma_k^2$,
so WSR is bounded by the finite sum
$\sum_k\omega_k\log_2(1+\|\tilde{\mathbf h}_{k,a_k}\|^2P_{a_k}^{\max}/\sigma_k^2)$.
Thus the objective values converge.
\end{proof}

\hypertarget{proof:D}{}\begin{proof}[{\proofref{prop:csi_dependence}{D. Channel blocks used by the updates}}]\leavevmode\\*
For $j\in\mathcal K_b$, $\mathbf w_{j,b}$ is supported on
$\mathcal V_{j,b}\subseteq\mathcal U_b$, while $\tilde{\mathbf h}_{k,b}$ is
supported on $\mathcal V_{k,b}$. Receiver and weight updates thus use only
$\mathcal V_{k,b}\cap\mathcal V_{j,b}\subseteq\mathcal V_{k,b}\cap\mathcal U_b$.
The beam update for $i\in\mathcal K_b$ uses $\tilde{\mathbf h}_{j,b,\mathcal V_i}$
only for $j\in\mathcal N_b(i)$, hence only
$\mathcal V_{j,b}\cap\mathcal V_{i,b}\subseteq\mathcal V_{j,b}\cap\mathcal U_b$.
The multiplier bisection uses only beamformer norms. These are precisely
\eqref{eq:conflict_blocks}.
\end{proof}

\hypertarget{proof:E}{}\begin{proof}[{\proofref{cor:conflict_load}{E. Conflict-aware upload load}}]\leavevmode\\*
Since $\mathcal V_{j,b}\cap\mathcal U_b\subseteq\mathcal V_{j,b}$,
termwise cardinalities give $\mathcal L_b^{\mathrm{conf}}\leq\mathcal L_b^{\mathrm{VR}}$.
Equality holds iff every difference is zero, equivalently
$\mathcal V_{j,b}\subseteq\mathcal U_b$ for all $j$.
\end{proof}

\hypertarget{proof:F}{}\begin{proof}[{\proofref{prop:decoupling}{F. Decoupling under empty effective victim sets}}]\leavevmode\\*
If $b\ne a_k$ and $j\in\mathcal K_b$ had
$\mathcal V_{k,b}\cap\mathcal V_{j,b}\ne\emptyset$, then $k$ would be an
effective victim, a contradiction. Hence all inter-cell terms vanish by
\hyperref[lem:interference]{Lemma~\ref*{lem:interference}}, proving (i). An outside-cell $j$ in
$\mathcal N_b(i)$ for $i\in\mathcal K_b$ would likewise be an effective victim;
thus $\mathcal N_b(i)\subseteq\mathcal K_b$. With the synchronized stopping schedule stated in (ii), the per-cell and multi-cell updates, initializations, WSR values and safeguards are identical by induction,
proving (ii).
\end{proof}

\hypertarget{proof:G}{}\begin{proof}[{\proofref{prop:expected_loads}{G. Expected VR-aware and conflict-aware loads}}]\leavevmode\\*
Linearity applied to \eqref{eq:csi_vr} gives
$\mathbb E[\mathcal L_b^{\mathrm{VR}}]=N_{\mathrm s}\sum_{k,g}q_{k,b,g}$
without independence. For conflict load, its indicator is
$\mathbf1\{g\in\mathcal V_{j,b}\}\mathbf1\{g\in\mathcal U_b\}$.
For $j\in\mathcal K_b$ the product is the first indicator. Otherwise,
\hyperlink{assumption:S1}{(S1)}--\hyperlink{assumption:S2}{(S2)} give expectation $q_{j,b,g}\Pr(g\in\mathcal U_b)$, where
$\Pr(g\in\mathcal U_b)=1-\prod_{i\in\mathcal K_b}(1-q_{i,b,g})=\zeta_{b,g}$.
Summing proves \eqref{eq:exp_load_conf}; \hyperlink{assumption:S3}{(S3)} is unrestricted.
\end{proof}

\hypertarget{proof:H}{}\begin{proof}[{\proofref{lem:interf}{H. True interference, per term}}]\leavevmode\\*
Decompose $\mathbf h_{k,b}=\tilde{\mathbf h}_{k,b}+\mathbf r_{k,b}$.
On the support $\mathcal V_{j,b}$ of $\mathbf w_j$, the residual is supported
only on $\mathcal V_{j,b}\setminus\mathcal V_{k,b}$; Cauchy--Schwarz gives
the bound. For disjoint VRs the gated term is zero and residual equals the
true channel on $\mathcal V_{j,b}$.
\end{proof}

\hypertarget{proof:I}{}\begin{proof}[{\proofref{thm:gap}{I. Model-to-true SINR and rate gap}}]\leavevmode\\*
\hyperref[lem:desired]{Lemma~\ref*{lem:desired}} makes $S_k$ exact. \hyperref[lem:interf]{Lemma~\ref*{lem:interf}} and
Minkowski's inequality give
$I_k^{\rm true}\leq(\sqrt{\hat I_k}+\sqrt{\Delta_k})^2$ and
$I_k^{\rm true}\geq(\sqrt{\hat I_k}-\sqrt{\Delta_k})_+^2$, proving (a).
For (b), set $D_k=\hat I_k+\sigma_k^2$ and
$E_k=(\sqrt{\hat I_k}+\sqrt{\Delta_k})^2+\sigma_k^2$. Since $D_k\leq E_k$,
\[
\hat R_k-R_k^{\mathrm{true}}\leq\log_2\frac{1+S_k/D_k}{1+S_k/E_k}
\leq\log_2(E_k/D_k)=B_k.
\]
The reverse-interference inequality gives the lower rate bound. Residual
interference can cancel gated interference, so the gap has no fixed sign.
Taylor expansion of $B_k$ gives the stated order. For (c), gated inter-cell
terms vanish; support, Cauchy--Schwarz and the per-BS power constraint give
$I_{k,b}^{\rm true}\leq\|\mathbf h_{k,b}[\mathcal U_b]\|^2P_b^{\max}$.
The disjointness hypothesis makes this restriction a part of $\mathbf r_{k,b}$,
which proves the stated $\epsilon_{k,b}$ bound.
\end{proof}

\hypertarget{proof:J}{}\begin{proof}[{\proofref{thm:proj}{J. Support-constraint optimality gap}}]\leavevmode\\*
Each projection $\mathbf P_j$ preserves VR support and contracts power, so
$\mathbf P\mathbf W^\star$ is feasible. Cauchy--Schwarz gives
$|\mathbf h_{k,a_k}^{\mathrm H}(\mathbf I-\mathbf P_k)\mathbf w_k^\star|\leq\sqrt{\Delta_k^\epsilon}$.
The reverse triangle inequality gives
$\sqrt{S_k'}\geq(\sqrt{S_k^\star}-\sqrt{\Delta_k^\epsilon})_+$,
which proves the numerator bound. It also yields
$S_k'\geq S_k^\star-2\sqrt{S_k^\star\Delta_k^\epsilon}$.
For each interferer, the projection error is at most
$\|\mathbf h_{k,b}[\mathcal V_{j,b}^c]\|\sqrt{\psi_j}\|\mathbf w_j^\star\|$;
Minkowski gives $I_k'\leq(\sqrt{I_k^\star}+\sqrt{\Delta_k^\psi})^2$.
With $D_k^\star=I_k^\star+\sigma_k^2$ and
$D_k'=(\sqrt{I_k^\star}+\sqrt{\Delta_k^\psi})^2+\sigma_k^2$,
\[
R_k^\star-R_k'\leq-\log_2(1-x_k)+\log_2(D_k'/D_k^\star),
\]
using $2\sqrt{S_k^\star\Delta_k^\epsilon}/(S_k^\star+D_k^\star)\leq x_k$.
Substitution gives $G$ after weighting and summing. The bound is vacuous
for $x_k\geq1$; both perturbations vanish as all $\psi_j\to0$.
\end{proof}

\hypertarget{proof:K}{}\begin{proof}[{\proofref{prop:fresnel}{K. Diffraction resolution}}]\leavevmode\\*
(a) Similar triangles give $h=y(d-x)/d$. The paraxial path expansion gives
$D_1+D_2-r=h^2d/[2x(d-x)]=\lambda v^2/4$, hence
$v=y\sqrt{2(d-x)/(\lambda xd)}$ and
$F=\arctan(\pm(\pi^{3/2}/4)v)/\pi$. Inverting the loss gives $v_L$;
monotonicity in (a) makes $v_1,v_\eta$ the transition endpoints, whose
separation divided by $dv/dy$ gives $w_F$. (b) Block centers on opposite
sides are separated by more than $w_F$; their end insets are each at least
$L_{\rm block}/2$, so $D_{\rm ap}>w_F+L_{\rm block}$. Requiring both whole blocks to lie outside the transition
adds another $L_{\rm block}$.
\end{proof}

\IEEEtriggeratref{16}


\begin{thebibliography}{99}

\bibitem{3gpp38901}
3GPP, \href{https://www.etsi.org/deliver/etsi_tr/138900_138999/138901/19.04.00_60/tr_138901v190400p.pdf}{``Study on channel model for frequencies from 0.5 to
100 GHz,''} \href{https://www.etsi.org/deliver/etsi_tr/138900_138999/138901/19.04.00_60/tr_138901v190400p.pdf}{TR 38.901}, V19.4.0, Release 19, Jul. 2026.

\bibitem{xu2026}
H. Xu et al., \href{https://doi.org/10.1109/JSAC.2026.3650888}{``Near-Field Propagation and
Spatial Non-Stationarity Channel Model for 6-24 GHz (FR3)
Extremely Large-Scale MIMO: Adopted by 3GPP for 6G,''} IEEE J.
Sel. Areas Commun., vol.~44, pp.~3201--3218, 2026, doi:
\href{https://doi.org/10.1109/JSAC.2026.3650888}{10.1109/JSAC.2026.3650888} (arXiv:2506.17887).

\bibitem{wang2025edof}
Z. Wang et al., \href{https://doi.org/10.1109/TWC.2025.3531418}{``Analytical framework for effective degrees of freedom in near-field XL-MIMO,''} \emph{IEEE Trans. Wireless Commun.}, vol.~24, no.~4, pp.~3465--3482, 2025, doi: \href{https://doi.org/10.1109/TWC.2025.3531418}{10.1109/TWC.2025.3531418}.

\bibitem{xu2026thz}
H. Xu et al., \href{https://doi.org/10.1109/TWC.2025.3610962}{``Empirical study on near-field and spatial non-stationarity modeling for THz XL-MIMO channel in indoor scenario,''} \emph{IEEE Trans. Wireless Commun.}, vol.~25, pp.~4435--4451, 2026, doi: \href{https://doi.org/10.1109/TWC.2025.3610962}{10.1109/TWC.2025.3610962}.

\bibitem{jin2025gbsm}
Y. Jin, R. He, B. Ai, Y. Yuan, Y. Niu, and H. Zhang, \href{https://doi.org/10.1109/TVT.2025.3529866}{``Geometry-based stochastic MIMO channel model for near-field and far-field scenarios of integrated sensing and communications,''} \emph{IEEE Trans. Veh. Technol.}, vol.~74, no.~5, pp.~6928--6940, 2025, doi: \href{https://doi.org/10.1109/TVT.2025.3529866}{10.1109/TVT.2025.3529866}.

\bibitem{zhi2024vr}
K. Zhi et al., \href{https://doi.org/10.1109/JSAC.2024.3389128}{``Performance analysis and
low-complexity design for XL-MIMO with near-field
spatial non-stationarities,''} \emph{IEEE J. Sel. Areas
Commun.}, vol.~42, no.~6, pp.~1656--1672, 2024,
doi: \href{https://doi.org/10.1109/JSAC.2024.3389128}{10.1109/JSAC.2024.3389128}.

\bibitem{jayasinghe2025vr}
M. A. Jayasinghe, D. H. Gunasinghe, and G. Amarasuriya,
\href{https://doi.org/10.1109/GLOBECOM59602.2025.11432685}{``Impact of visibility regions on multi-cell MIMO NOMA
systems with extra-large antenna arrays,''} in \emph{Proc.
IEEE GLOBECOM}, 2025, pp.~3630--3635,
doi: \href{https://doi.org/10.1109/GLOBECOM59602.2025.11432685}{10.1109/GLOBECOM59602.2025.11432685}.

\bibitem{perezadan2025}
D. P\'erez-Ad\'an, J. P. Gonz\'alez-Coma, F. J. L\'opez-Mart\'inez, and
L. Castedo, \href{https://doi.org/10.1109/LWC.2025.3543348}{``Interference-Aware Precoding and User Selection
for Multi-Cell Near-Field XL-MIMO Systems,''} IEEE Wireless
Commun. Lett., vol.~14, no.~5, pp.~1396--1400, 2025,
doi: \href{https://doi.org/10.1109/LWC.2025.3543348}{10.1109/LWC.2025.3543348}.

\bibitem{xing2025wn}
H. Xing, Y. Zhang, J. Zhang, H. Xu, G. Liu, and Q. Wang, \href{https://doi.org/10.1109/TVT.2025.3531041}{``An approximate wave-number domain expression for near-field XL-array channel,''} \emph{IEEE Trans. Veh. Technol.}, vol.~74, no.~5, pp.~8267--8272, 2025, doi: \href{https://doi.org/10.1109/TVT.2025.3531041}{10.1109/TVT.2025.3531041}.

\bibitem{zheng2025elaa}
F. Zheng, H. Yu, C. Wang, L. Sun, Q. Wu, and Y. Chen, \href{https://doi.org/10.1109/TCOMM.2025.3535864}{``Extremely large-scale array systems: Near-field codebook design and performance analysis,''} \emph{IEEE Trans. Commun.}, vol.~73, no.~8, pp.~6798--6812, 2025, doi: \href{https://doi.org/10.1109/TCOMM.2025.3535864}{10.1109/TCOMM.2025.3535864}.

\bibitem{li2025codebook}
X. Li, Z. Dong, Y. Zeng, and Y. Li, \href{https://doi.org/10.1109/TCOMM.2025.3577652}{``Codebook design and beam training for multi-user modular XL-MIMO communications: From far-field to near-field,''} \emph{IEEE Trans. Commun.}, vol.~73, no.~11, pp.~11855--11869, 2025, doi: \href{https://doi.org/10.1109/TCOMM.2025.3577652}{10.1109/TCOMM.2025.3577652}.

\bibitem{meng2025isac}
C. Meng, D. Ma, Z. Wang, Y. Liu, Z. Wei, and Z. Feng, \href{https://doi.org/10.1109/TCOMM.2025.3573460}{``Near-field hybrid beamforming design for modular XL-MIMO ISAC systems,''} \emph{IEEE Trans. Commun.}, vol.~73, no.~11, pp.~11840--11854, 2025, doi: \href{https://doi.org/10.1109/TCOMM.2025.3573460}{10.1109/TCOMM.2025.3573460}.

\bibitem{lu2025flex}
H. Lu, H. Min, Y. Zeng, and S. Ma, \href{https://doi.org/10.1109/TCOMM.2025.3622989}{``Flexible XL-MIMO via array configuration codebook: Codebook design and array configuration training,''} \emph{IEEE Trans. Commun.}, vol.~73, pp.~15190--15204, 2025, doi: \href{https://doi.org/10.1109/TCOMM.2025.3622989}{10.1109/TCOMM.2025.3622989}.

\bibitem{lu2026compact}
J. Lu et al., \href{https://doi.org/10.1109/TCOMM.2025.3649926}{``Compact ultra massive antenna arrays under mutual coupling: Modeling and spectral efficiency analysis,''} \emph{IEEE Trans. Commun.}, vol.~74, pp.~3174--3189, 2026, doi: \href{https://doi.org/10.1109/TCOMM.2025.3649926}{10.1109/TCOMM.2025.3649926}.

\bibitem{cao2025ris}
X. Cao, M. Mohammadi, H. Q. Ngo, H. Shin, and M. Matthaiou, \href{https://doi.org/10.1109/TWC.2025.3617656}{``RIS-assisted XL-MIMO for near-field and far-field communications,''} \emph{IEEE Trans. Wireless Commun.}, vol.~25, pp.~5320--5338, 2026, doi: \href{https://doi.org/10.1109/TWC.2025.3617656}{10.1109/TWC.2025.3617656}.

\bibitem{che2025pilot}
H. Che, L. You, J. Wang, and X. Gao, \href{https://doi.org/10.1109/TVT.2024.3467113}{``Non-orthogonal pilot design exploiting spatial non-stationarity for TDD XL-MIMO transmission,''} \emph{IEEE Trans. Veh. Technol.}, vol.~74, no.~2, pp.~3424--3428, 2025, doi: \href{https://doi.org/10.1109/TVT.2024.3467113}{10.1109/TVT.2024.3467113}.

\bibitem{zhang2026pilot}
Y. Zhang, H. Guo, and V. K. N. Lau, \href{https://doi.org/10.1109/TSP.2026.3661556}{``A novel pilot scheme for uplink channel estimation for sub-array structured ELAA in XL-MIMO systems,''} \emph{IEEE Trans. Signal Process.}, vol.~74, pp.~2787--2802, 2026, doi: \href{https://doi.org/10.1109/TSP.2026.3661556}{10.1109/TSP.2026.3661556}.

\bibitem{yilmaz2026vr}
Y. Yilmaz, S. Demirci, L. Afeef, and H. Arslan,
\href{https://doi.org/10.1109/TCOMM.2026.3692106}{``VR-based spatial and temporal domains interference
suppression in XL-MIMO systems,''} \emph{IEEE Trans.
Commun.}, vol.~74, pp.~8718--8734, 2026,
doi: \href{https://doi.org/10.1109/TCOMM.2026.3692106}{10.1109/TCOMM.2026.3692106}.

\bibitem{shi2011}
Q. Shi, M. Razaviyayn, Z.-Q. Luo, and C. He, \href{https://doi.org/10.1109/TSP.2011.2147784}{``An iteratively
weighted MMSE approach to distributed sum-utility maximization
for a MIMO interfering broadcast channel,''} IEEE Trans. Signal
Process., vol.~59, no.~9, pp.~4331--4340, Sep. 2011,
doi: \href{https://doi.org/10.1109/TSP.2011.2147784}{10.1109/TSP.2011.2147784}.

\bibitem{wang2026cf}
Z. Wang et al., \href{https://doi.org/10.1109/TWC.2026.3659809}{``Low-complexity distributed combining design for near-field cell-free XL-MIMO systems,''} \emph{IEEE Trans. Wireless Commun.}, vol.~25, pp.~11799--11815, 2026, doi: \href{https://doi.org/10.1109/TWC.2026.3659809}{10.1109/TWC.2026.3659809}.

\bibitem{liao2026edge}
X.~Liao et al., \href{https://doi.org/10.1109/TWC.2026.3660301}{``Edge-enhanced distributed
downlink power control and UE association in scalable
cell-free massive MIMO systems,''} \emph{IEEE Trans.
Wireless Commun.}, vol.~25, pp.~11383--11398, 2026.
doi: \href{https://doi.org/10.1109/TWC.2026.3660301}{10.1109/TWC.2026.3660301}.

\bibitem{chen2025meta}
Y.~Chen et al.,
\href{https://doi.org/10.1109/TWC.2025.3618804}{``Meta-hierarchical reinforcement learning-based
beamforming for near-field multi-user communications,''}
\emph{IEEE Trans. Wireless Commun.}, vol.~25,
pp.~5446--5462, 2026 (early access, 2025).
doi: \href{https://doi.org/10.1109/TWC.2025.3618804}{10.1109/TWC.2025.3618804}.

\bibitem{cui2025ice}
M. Cui, Z. Zhang, L. Dai, and K. Huang, \href{https://doi.org/10.1109/TWC.2025.3567797}{``Ice-filling: Near-optimal channel estimation for dense array systems,''} \emph{IEEE Trans. Wireless Commun.}, vol.~24, no.~10, pp.~8551--8564, 2025, doi: \href{https://doi.org/10.1109/TWC.2025.3567797}{10.1109/TWC.2025.3567797}.

\bibitem{zhang2026obs}
Z. Zhang and M. Cui, \href{https://doi.org/10.1109/TSP.2026.3657342}{``Observation matrix design for densifying MIMO channel estimation via 2D ice filling,''} \emph{IEEE Trans. Signal Process.}, vol.~74, pp.~798--814, 2026, doi: \href{https://doi.org/10.1109/TSP.2026.3657342}{10.1109/TSP.2026.3657342}.

\bibitem{lu2025extrap}
J.~Lu, J.~Zhang, H.~Lei, H.~Xiao, B.~Ai, and
D.~W.~K. Ng, \href{https://doi.org/10.1109/TWC.2025.3610388}{``Near-field spatial-domain channel
extrapolation for XL-MIMO systems,''} \emph{IEEE Trans.
Wireless Commun.}, vol.~25, pp.~4343--4357, 2026.
doi: \href{https://doi.org/10.1109/TWC.2025.3610388}{10.1109/TWC.2025.3610388}.

\bibitem{mobini2026fh}
Z. Mobini, A. H. Gokceoglu, L. Wang, G. Peters, and H. Q. Ngo, \href{https://doi.org/10.1109/TWC.2025.3587672}{``Fronthaul-aware user-centric generalized cell-free massive MIMO systems,''} \emph{IEEE Trans. Wireless Commun.}, vol.~25, pp.~900--914, 2026, doi: \href{https://doi.org/10.1109/TWC.2025.3587672}{10.1109/TWC.2025.3587672}.

\bibitem{mobini2026cluster}
Z. Mobini, A. H. Gokceoglu, L. Wang, G. Peters, H. Shin, and H. Q. Ngo, \href{https://doi.org/10.1109/TWC.2025.3624199}{``Cluster-wise processing in fronthaul-aware cell-free massive MIMO systems,''} \emph{IEEE Trans. Wireless Commun.}, vol.~25, pp.~6240--6254, 2026, doi: \href{https://doi.org/10.1109/TWC.2025.3624199}{10.1109/TWC.2025.3624199}.

\end{thebibliography}
\end{document}